\documentclass[journal,draftcls,onecolumn,12pt,twoside]{IEEEtran}

\usepackage{cite}
\usepackage{amsmath,amssymb}
\usepackage{interval}
\usepackage{cool}
\usepackage{breqn}
\usepackage{wasysym}
\usepackage{dsfont}
\usepackage{graphicx}
\usepackage{times, epsfig}
\usepackage{subfig}
\usepackage{verbatim}
\usepackage[latin1]{inputenc}
\newtheorem{thm}{Theorem}

\newtheorem{defn}{Definition}

\newtheorem{rk}{Remark}
\newtheorem{cor}{Corollary}

\usepackage[usenames, dvipsnames]{color}
\usepackage{stfloats}

\graphicspath{ {./images/} }
\newlength{\figwidth}
\begin{document}
	\title{A Sample-Deficient Analysis of the Leading Generalized Eigenvalue for the Detection of Signals in Colored Gaussian Noise\vspace{-4mm}}

\author{
	\vspace*{-0mm}
	Prathapasinghe~Dharmawansa,~Saman~Atapattu,~\IEEEmembership{Senior Member, IEEE},\\ Jamie~Evans,~\IEEEmembership{Senior Member, IEEE}, and Kandeepan~Sithamparanathan,~\IEEEmembership{Senior~Member, IEEE} 
	\vspace*{-0mm}
	\thanks{This work was presented in part at the IEEE  Global Communications Conference  (GLOBECOM 23), Kuala Lumpur, Malaysia, Dec. 2023 \cite{Dharmawansa2023Globecom}.}
 \thanks{P. Dharmawansa and J. Evans are with the Department of Electrical and Electronic Engineering, University of Melbourne, Parkville, VIC 3010, Australia (e-mail: \{prathapa.d, jse\}@unimelb.edu.au).}%
		\thanks{S. Atapattu and K. Sithamparanathan are with the School of Engineering, RMIT University, Melbourne, VIC, Australia (e-mail: \{saman.atapattu, kandeepan.sithamparanathan\}@rmit.edu.au).}

\vspace*{-0mm}
}
 

	\maketitle
	
	

	%
    
    \begin{abstract}
This paper investigates the signal detection problem in colored Gaussian noise with an unknown covariance matrix. To be specific, we consider a sample deficient scenario in which the number of signal bearing samples ($n$) is strictly smaller than the dimensionality of the signal space ($m$). Our test statistic is the leading generalized eigenvalue of the whitened sample covariance matrix (a.k.a. $F$-matrix) which is constructed by whitening the signal bearing sample covariance matrix with noise-only sample covariance matrix. The whitening operation along with the observation model induces a single spiked covariance structure on the $F$-matrix. Moreover, 
the sample deficiency (i.e., $m>n$) in turn makes this $F$-matrix rank deficient, thereby {\it singular}.
Therefore, a simple exact statistical characterization of the leading generalized eigenvalue (l.g.e.) of a complex correlated {\it singular} $F$-matrix with a single spiked associated covariance is of paramount importance to assess the performance of the detector (i.e., the receiver operating characteristics (ROC)). To this end, we adopt the powerful orthogonal polynomial technique in random matrix theory to derive a new finite dimensional c.d.f. expression for the l.g.e. of this particular $F$-matrix. It turns out that when the noise only sample covariance matrix is nearly rank deficient and the signal-to-noise ratio is $O(m)$, the ROC profile converges to a remarkably simple limiting profile. 
    \end{abstract}
     \begin{IEEEkeywords}
     Colored noise, Detection, Eigenvalues, $F$-matrix, orthogonal polynomials, Random matrix, Receiver operating characteristics (ROC), singular Wishart matrix, Stiefel manifold
     \end{IEEEkeywords}

	\IEEEpeerreviewmaketitle

\vspace{-0.5cm}
\section{Introduction}

The detection of signals embedded in noise is a fundamental problem with numerous applications in various scientific disciplines \cite{Nadakuditi2008sp,Nadakuditi2010jsacsp,ONATSKI2014rmt,chamain2020eigenvalue,Johnstone2017Biometrika,dogandzic2003generalized, kritchman2009non,bianchi2011performance,he2010mimo,mcwhorter2023first}. In this respect, the test statistic based on the leading eigenvalue of the sample covariance matrix--also known as Roy's largest root\footnote{This is a direct consequence of Roy's union-intersection principle \cite{Mardia1979book}.}--has been popular among detection theorists \cite{ONATSKI2014rmt,chamain2020eigenvalue,Johnstone2017Biometrika,dharmawansa2019roy,Dharmawansa2014arx2,wang2017stat,namdari2021high,kritchman2009non}. This stems from the fact that the largest
root test is most powerful among the common tests when
the alternative is of rank-one \cite{Johnstone2017Biometrika,kritchman2009non,memuirhead2009aspects}.
In its most basic form with Gaussian signal and additive white Gaussian noise assumptions, this amounts to statistically characterizing the largest eigenvalue of a Wishart matrix having a so-called {\it spiked covariance} structure, see e.g., \cite{me2,Johnstone2017Biometrika,dharmawansa2019roy, ref:couillet} and references therein. The spiked covariance structure that deviates from a benchmark matrix model, usually the identity matrix, along a small number of arbitrary directions (or spikes) turns out to be an abstraction of considerable importance in various research disciplines. For instance, when the number of directions is one (i.e., rank-one/single spiked model), certain optimal properties of the leading sample eigenvalue, as a test statistic, in the asymptotic sample regime has been established in \cite{kritchman2009non}.   

The white Gaussian noise assumption, though very common in the classical setting, may not hold in certain practical scenarios \cite{Maris2003biomed,Vinogradova2013sp,Hiltunen2015sp,Nadakuditi2017it,dogandzic2003generalized,werner2007doa,werner2006optimal,ding2022tracy,leeb2021optimal,he2010mimo,couillet2012fluctuations,bautista2019detecting,agterberg2022entrywise,mcwhorter2023first,vallet2017performance,Cohen,Kliger,Navarro,viswanath2002optimal,Tam,Chong,Uzi,YCliang2023,Lozano2003multiple}. To be specific, many applications in signal processing, wireless communications, RADAR, imaging, and related
fields involve noise that is colored\footnote{Here we refer to spatially colored Gaussian noise which is temporally white. The detection in temporally colored Gaussian noise has been treated in detail in \cite{kailath1,kailath2,kailath3,vantreedetection}.} with unknown covariance structure.
In such situations, for decision theoretic purposes, the generalized eigenvalues of the so-called whitened signal-plus-noise {\it sample} covariance matrix (a.k.a. $F$-matrix) have been employed \cite{Nadakuditi2010jsacsp, dharmawansa2019roy, chamain2020eigenvalue,Johnstone2017Biometrika,werner2007doa,werner2006optimal,Gavish2023,leeb2021optimal,landa2022biwhitening,tani2023efficient}. The whitening operation entails rotation and rescaling the coordinates, after which the effective noise becomes white.
In particular, the whitening operation requires two sample covariance matrices: noise only (i.e., signal free) denoted by $\widehat{\boldsymbol{\Sigma}}_n$ and signal-plus-noise denoted by $\widehat{\boldsymbol{\Sigma}}_s$ \cite{Nadakuditi2010jsacsp,chamain2020eigenvalue,Johnstone2017Biometrika,dharmawansa2019roy,werner2007doa,werner2006optimal,dogandzic2003generalized,mcwhorter2023first,vallet2017performance}. In this regard, the noise-only sample covariance matrix can be formed in many domain-specific scenarios as delineated in  \cite{Nadakuditi2010jsacsp,werner2007doa,werner2006optimal,dogandzic2003generalized,Gavish2023,Zhigang2015,mariani2019oversampling,he2010mimo,richmond2006mean,zachariah2013utilization}. To be precise, assume that we have $n$ possible i.i.d. signal bearing signal-plus-noise sample observations\footnote{This is also known as the primary data set \cite{richmond2006mean}.} denoted by $\{\mathbf{x}_1, \mathbf{x}_2,\ldots,\mathbf{x}_n\}$, where $\mathbf{x}_k\in\mathbb{C}^{m\times 1}$, and $p$ i.i.d. {\it noise only} sample observations\footnote{This is also known as the secondary data set \cite{richmond2006mean}.} denoted by $\{\mathbf{n}_1,\mathbf{n}_2,\ldots,\mathbf{n}_p\}$, where $\mathbf{n}_\ell\in\mathbb{C}^{m\times 1}$ at our disposal. Then we may express the corresponding sample covariance matrices as
\begin{align*}
	\widehat{\boldsymbol{\Sigma}}_s=\frac{1}{n}\sum_{k=1}^n \mathbf{x}_k\mathbf{x}_k^\dagger
	& \qquad \text{and}\qquad 
	\widehat{\boldsymbol{\Sigma}}_n=\frac{1}{p}\sum_{\ell=1}^p \mathbf{n}_\ell \mathbf{n}_\ell^\dagger
	\end{align*}
 where $(\cdot)^\dagger$ denotes the Hermitian transpose operator.
Consequently, the whitened signal-plus-noise sample covariance matrix takes the form $\widehat{\boldsymbol{\Sigma}}_n^{-1/2}\widehat{\boldsymbol{\Sigma}}_s\widehat{\boldsymbol{\Sigma}}_n^{-1/2}$, where $\widehat{\boldsymbol{\Sigma}}_n^{-1/2}$ denotes the inverse of the symmetric positive definite square root of the symmetric positive definite matrix $\widehat{\boldsymbol{\Sigma}}_n$. 

To be consistent, one has to make sure that the number of noise only samples $p$ is greater than or equal to the dimensionality  of the system $m$ (i.e., $p\geq m$) so that the noise-only sample covariance matrix $\widehat{\boldsymbol{\Sigma}}_n$ is invertible almost surely \cite{memuirhead2009aspects}. As for the number of signal-plus-noise samples $n$, it is common to make the classical assumption that $n\geq m$  to ensure the almost sure positive definiteness of $\widehat{\boldsymbol{\Sigma}}_s$ \cite{memuirhead2009aspects}. However, the $n<m$ scenario (i.e., sample deficiency/high dimensionality) is becoming increasingly common in modern applications \cite{Nadakuditi2008sp,Vinogradova2013sp,vallet2017performance,pham2015performance,vallet2015performance,mestre2008modified}. For instance, as expounded in \cite{Vinogradova2013sp,vallet2017performance,vallet2015performance,pham2015performance}, on account of the increased antenna array size and the urge for faster decisions (i.e., detection and estimation dynamics), the operational regime of  modern array processing techniques has shifted to the low sample regime. 
In this sample deficient regime, the signal-plus-noise sample covariance matrix $\widehat{\boldsymbol{\Sigma}}_s$ becomes rank deficient (i.e., singular) almost surely \cite{srivastava2003singular, mallik2003pseudo,uhlig1994singular,ratnarajah2005complex, Onatski2008Ann}, thereby serving as a poor estimator of the population signal-plus-noise covariance matrix. This in turn makes the whitened signal-plus-noise sample covariance matrix also singular. In addition to the rank deficiency, the whitening operation along with the specific sampling/observational model  induces a single spiked covariance structure on $\widehat{\boldsymbol{\Sigma}}_n^{-1/2}\widehat{\boldsymbol{\Sigma}}_s\widehat{\boldsymbol{\Sigma}}_n^{-1/2}$.
As we shall see below, the joint effect of sample deficiency driven singularity and induced covariance pose major technical challenges in finite dimensional statistical characterization of the largest root of the complex correlated singular $F$-matrix $\widehat{\boldsymbol{\Sigma}}_n^{-1/2}\widehat{\boldsymbol{\Sigma}}_s\widehat{\boldsymbol{\Sigma}}_n^{-1/2}$.

The  fundamental asymptotic statistical characterizations, in sample size, dimensionality, and signal-to-noise-ratio (SNR), of the largest generalized root of the correlated $F$-matrix for $n\geq m$ scenario have been thoroughly studied in the literature \cite{Nadakuditi2010jsacsp,Dharmawansa2014arx2,Johnstone2017Biometrika,Johnstone2008stat,jiang2013Bernoulli,Dumitriu2008siam,Johnstone2020stat,Zhigang2015,Gavish2023,forrester2010log,meJames,wang2017extreme,chiani2017probability,chiani2016distribution}. The finite dimensional characteristics corresponding to the same scenario have  been analyzed in  \cite{chamain2020eigenvalue,Dumitriu2008siam,chiani2017probability,chiani2016distribution,dubbs2014beta,kan2019densities}. Nevertheless, to our best knowledge, a tractable {\it finite dimensional} analysis for $n<m$ (i.e., sample deficient) scenario is not available in the literature. Thus,  in this paper, we focus on this sample deficient regime.

Under the Gaussian assumption with $n<m$, the largest generalized sample eigenvalue based detection in colored noise  amounts to finite dimensional statistical characterization of the largest eigenvalue of  complex correlated {singular} $F$-matrix. In this respect, the joint eigenvalue density of the uncorrelated real {\it singular} $F$-matrix has been derived in \cite{srivastava2003singular,diaz1997proof,uhlig1994singular,shimizu2022expressing}. The joint eigenvalue density of
complex correlated {\it singular} $F$-matrix, which contains the so-called {\it heterogeneous} hypergometric function of two matrix arguments, has been reported in \cite{shimizu2022expressing}. An expression involving heterogeneous hypergometric function of one matrix argument for the largest generalized eigenvalue has also been derived therein. However, the algebraic complexity of these hypergemoetric functions  in turn makes them less amenable to further analysis.
 Therefore, in this paper, capitalizing on powerful contour integral approach due to \cite{meWang} and orthogonal polynomial techniques due to \cite{me11}, {\bf we derive simple and tractable closed-form expressions for the joint eigenvalue density and the cumulative distribution function (c.d.f.) of the maximum generalized eigenvalue of the complex correlated {\it singular} $F$-matrix when the underlying covariance matrix assumes a single spiked structure.} The resultant c.d.f. expression consists of a determinant of a square matrix whose dimensions depend on the relative difference between the number of noise only samples $p$ and the system dimensionality $m$ (i.e., $p-m$) but not their individual magnitudes. This key feature further enables us to bypass the determinant evaluation process in expressing the c.d.f. corresponding to an important configuration $p=m$. Since the parameter $p-m$ can also be used as an implicit indicator of the quality of $\widehat{\boldsymbol{\Sigma}}_n$ as an estimator of the unknown population noise covariance matrix, the above configuration corresponds to the lowest quality noise covariance estimator. Therefore, this configuration in turn dictates a performance lower bound on the leading eigenvalue as a test statistic. Apart from these developments, this new c.d.f. expression further facilitates the analysis of the receiver operating characteristics (ROC) of the largest root test in the sample deficient regime. 

The key results developed in this paper shed some light on the impact of the  the system dimension ($m$), the number of signal-plus-noise samples  ($n$) and noise-only   observations ($p$), and the SNR ($\gamma$) on the ROC in the $n<m$ regime.
For instance, our results reveal that the relative disparity between $m$ and $n$ degrades the detection power of the test for fixed values of the other parameters. This stems from the fact that the quality of the signal-plus-noise sample covariance estimate $\widehat{\boldsymbol{\Sigma}}_s$, which is already a poor estimator of the population signal-plus-noise covariance matrix, further degrades with increasing $m-n$. It turns out that, for fixed $n,\gamma$ with $p=m$ (i.e., when the noise-only sample covariance matrix is nearly rank deficient), the leading eigenvalue has no power in the asymptotic domain in which $m\to\infty$. The reason for this is not intuitive but rather technical. Following \cite{Nadakuditi2010jsacsp,Dharmawansa2014arx2,Johnstone2020stat}, it can be observed that, under the scaling $m=p$ with $m\to\infty$, the operation regime lies below the {\it phase transition}\footnote{This phenomenon is commonly known as the Baik, Ben Arous, P\'ech\'e (BBP) phase transition because of their seminal contribution in \cite{ref:baikPhaseTrans}. The signal processing analogy of this phenomenon is known as the ``subspace swap" \cite{ref:johnsonMestre},\cite{ref:thomas}, \cite{ref:tuft}.} (i.e., subcritical regime) in which the leading eigenvalue has no power.\footnote{To be precise, in this regime, properly centered and scaled leading eigenvalue converges in distribution to the well-celebrated Tracy-Widom distribution \cite{ref:tracy} in both situations; in the presence of a signal as well as in the absence \cite{Nadakuditi2010jsacsp,Dharmawansa2014arx2,Johnstone2020stat,wang2017stat}.}   
However, when $\gamma=O(m)$ and $p=m$, the ROC profile converges to a remarkably simple limiting profile  as $m\to\infty$, thereby the leading eigenvalue retaining its power asymptotically. Our numerical results further reveal that this particular limiting profile serves as a very good approximation to finite dimensional configurations as well. 
The utility of this insight in designing and analyzing future wireless signal processing techniques is brought into sharp focus by noting the fact that, under Rayleigh fading, as $m\to\infty$, the SNR $\gamma$ scales with $m$ almost surely due to the strong law of large numbers.

The remainder of this paper is organized as follows. In Section II, we formulate the signal detection problem in colored noise with an unknown noise covariance matrix as a binary hypothesis testing problem. Section III derives the novel c.d.f. expression for the leading generalized eigenvalue, which is the proposed test statistic for the preceding binary hypothesis testing problem, of a complex singular correlated $F$ matrix with a rank-one spiked  underlying covariance structure. This is made possible due to a new expression for the joint density of the eigenvalues of the same random matrix that we have derived in the beginning of Section III. The ROC performance of the leading generalized eigenvalue with respect to the system parameters (i.e., $m,n,p$ and $\gamma$) in a sample deficient scenario is analyzed in Section IV. Certain asymptotic results pertaining to the ROC corresponding to the important configuration $p=m$ are  also derived therein. Finally, conclusive remarks are made in Section V.

{\it Notation}: The following notation is used throughout this paper. 
A complex Gaussian random variable $X$ with zero mean and variance $\sigma^2$ is denoted by $X\sim \mathcal{CN}(0,\sigma^2)$. 
The superscript $(\cdot)^\dagger$ indicates the Hermitian transpose, $\Re\{\cdot\}$ denotes the real part of a complex number,  $\text{det}(\cdot)$ denotes the determinant of a square matrix, $\text{tr}(\cdot)$ represents the trace of a square matrix, and $\text{etr}(\cdot)$ stands for $\exp\left(\text{tr}(\cdot)\right)$. The $n\times n$ identity matrix is represented by $\mathbf{I}_n$ and the Euclidean norm of a vector $\mathbf{w}$ is denoted by $||\mathbf{w}||$. The symmetric positive definite square root of a symmetric positive definite matrix $\mathbf{B}$ is denoted by $\mathbf{B}^{1/2}$. A diagonal matrix with the diagonal entries $a_1,a_2,\ldots, a_n$ is denoted by $\text{diag}(a_1, a_2,\ldots,a_n)$. We denote the $m\times m$ unitary group by $\mathcal{U}_m$, whereas the set of all $m\times n$ ($m>n$) complex matrices $\mathbf{U}_1$ such that $\mathbf{U}_1^\dagger \mathbf{U}_1=\mathbf{I}_n$ (i.e., with orthonormal columns), denoted by $\mathcal{V}_{n,m}$, is known as the complex {\it Stiefel manifold}. Finally, we use the following notation to compactly represent the
determinant of an $n\times n$ block matrix:
\begin{equation*}
\begin{split}
\det\left[a_{i}\;\; b_{i,j}\right]_{\substack{i=1,2,\ldots,n\\
 j=2,3,\ldots,n}}&=\left|\begin{array}{ccccc}
 a_{1} & b_{1,2}& b_{1,3}& \ldots & b_{1,n}\\
  \vdots & \vdots & \vdots &\ddots & \vdots \\
  a_{n} & b_{n,2}& b_{n,3}& \ldots & b_{n,n}
 \end{array}\right|
 \end{split}
\end{equation*}
and $\det[a_{i,j}]_{i,j=1,\ldots,N}$ denotes the determinant of an $N\times N$ square matrix with its $(i,j)^{\text{th}}$ element given by $a_{i,j}$.

	\section{Problem formulation}
	
 	Consider the following general signal detection problem in colored Gaussian noise
\begin{align}
\mathbf{x}=\sqrt{\rho}\mathbf{h}s+\mathbf{n}\end{align}
	where $\mathbf{x}\in\mathbb{C}^{m}$, $\mathbf{h}\in\mathbb{C}^{m}$ is an unknown channel vector, $\rho\geq 0$, $s\sim\mathcal{CN}(0,1)$ is the signal, and  $\mathbf{n}\sim \mathcal{CN}_m(\mathbf{0}, \boldsymbol{\Sigma})$ denotes the colored noise which is independent of $s$. Moreover, the noise covariance matrix $\boldsymbol{\Sigma}\in\mathbb{C}^{m\times m}$ is unknown at the detector. Now the classical signal detection problem reduces to the following binary hypothesis testing problem
	\begin{align*}
	&\mathcal{H}_0:\; \rho=0\;\;\;\;\;\; \text{Signal is absent}\\
	& \mathcal{H}_1:\; \rho>0 \;\;\;\;\; \text{Signal is present}.
	\end{align*}
Noting that the covariance matrix of $\mathbf{x}$ assumes two different structures under the two hypotheses, the above testing problem can be written in terms of covariance matrices as
	\begin{align*}
	\begin{array}{ll}
	\mathcal{H}_0:\; \boldsymbol{\Sigma}_n=\boldsymbol{\Sigma} &\text{Signal is absent}\\
	\mathcal{H}_1:\; \boldsymbol{\Sigma}_s=\rho\mathbf{h}\mathbf{h}^\dagger+\boldsymbol{\Sigma} & \text{Signal is present}.
	\end{array}
	\end{align*}
	Let us now consider the whitened symmetric matrix \begin{align}
\boldsymbol{\Theta}=\boldsymbol{\Sigma}_n^{-1/2}\boldsymbol{\Sigma}_s \boldsymbol{\Sigma}_n^{-1/2}=\boldsymbol{\Sigma}^{-1/2}\mathbf{h}\mathbf{h}^\dagger \boldsymbol{\Sigma}^{-1/2}+\mathbf{I}_m
	\end{align} with the generalized eigenvalues $\lambda_1\leq \lambda_2\leq \ldots\leq \lambda_m$. Since $\mathbf{hh}^\dagger$ is a rank-$1$ matrix, we readily obtain $\lambda_m=1+\mathbf{h}^\dagger \boldsymbol{\Sigma}^{-1}\mathbf{h}>1$, whereas $\lambda_1=\lambda_2=\ldots=\lambda_{m-1}=1$. This  discrimination power of $\lambda_m$--the leading generalized eigenvalue-- indicates its utility as a test statistic in the above hypothesis testing problem \cite{Nadakuditi2010jsacsp,Johnstone2017Biometrika,dharmawansa2019roy,Dharmawansa2014arx2,chamain2020eigenvalue}.

	In most practical scenarios, the covariance matrices $\boldsymbol{\Sigma}_n$ and $\boldsymbol{\Sigma}_s$ are unknown so that the above procedure cannot be trivially applied. To circumvent this difficulty, the covariance matrices $\boldsymbol{\Sigma}_n$ and $\boldsymbol{\Sigma}_s$ are commonly replaced by their sample estimates. To be precise, let us assume that we have $n \geq 1 $ i.i.d. sample observations from signal-plus-noise scenario given by $\{\mathbf{x}_1, \mathbf{x}_2,\ldots, \mathbf{x}_n\}$ and $p>1$ i.i.d. sample observations from noise-only scenario\footnote{This is a commonly used  assumption in the literature, see e.g., \cite{Nadakuditi2010jsacsp,Zhigang2015,mariani2019oversampling,werner2007doa,werner2006optimal,dogandzic2003generalized,Gavish2023, Johnstone2017Biometrika,richmond2006mean,zachariah2013utilization}.}
given by $\{\mathbf{n}_1, \mathbf{n}_2,\ldots,\mathbf{n}_p\}$. Consequently,  the  sample estimates of $\boldsymbol{\Sigma}_n$ and $\boldsymbol{\Sigma}_s$ become 
	\begin{align}
	\widehat{\boldsymbol{\Sigma}}_n=\frac{1}{p}\sum_{\ell=1}^p \mathbf{n}_\ell \mathbf{n}_\ell^\dagger
	\quad \text{and}\quad
	\widehat{\boldsymbol{\Sigma}}_s=\frac{1}{n}\sum_{k=1}^n \mathbf{x}_k\mathbf{x}_k^\dagger.
	\end{align} 
Here we assume that the number of noise only samples is at least the dimensionality of the system (i.e., $p\geq m$), whereas the number of possible signal-plus-noise samples is strictly smaller than the dimensionality of the system (i.e., $m>n$). The latter assumption in turn  makes the estimated covariance matrix $\widehat{\boldsymbol{\Sigma}}_s$ rank deficient (i.e., rank at most $n$) and therefore, {\it singular}. Consequently, following \cite{Nadakuditi2010jsacsp,Johnstone2017Biometrika,dharmawansa2019roy,Dharmawansa2014arx2}, we form the {\it singular} matrix
	\begin{align}
\widehat{\boldsymbol{\Theta}}=\widehat{\boldsymbol{\Sigma}}_n^{-1/2}\widehat{\boldsymbol{\Sigma}}_s\widehat{\boldsymbol{\Sigma}}_n^{-1/2}
	\end{align}
	and investigate its maximum eigenvalue as the test statistic. Since $\widehat{\boldsymbol{\Theta}}$ is rank $n$ Hermitian positive semi-definite matrix, its ordered non-zero eigenvalues can be written as $0<\hat{\lambda}_{1}<\hat{\lambda}_{2}<\ldots<\hat{\lambda}_{n}$.
 In this respect, from distributional point of view, we have $
p\widehat{\boldsymbol{\Sigma}}_n\sim\mathcal{CW}_m\left(p, \boldsymbol{\Sigma}\right)$ and most importantly, $n\widehat{\boldsymbol{\Sigma}}_s$ assumes a {\it singular Wishart} density (i.e., due to $m>n$) given by $
n\widehat{\boldsymbol{\Sigma}}_s\sim\mathcal{SCW}_m\left(n, \boldsymbol{\Sigma}+\rho \mathbf{h}\mathbf{h}^\dagger\right)
	$.
	Keeping in mind that the eigenvalues of $ \widehat{\boldsymbol{\Theta}}$ do not change under the simultaneous transformations $\widehat{\boldsymbol{\Sigma}}_n\mapsto \boldsymbol{\Sigma}^{-1/2}\widehat{\boldsymbol{\Sigma}}_n\boldsymbol{\Sigma}^{-1/2}$, and $\widehat{\boldsymbol{\Sigma}}_s\mapsto \boldsymbol{\Sigma}^{-1/2}\widehat{\boldsymbol{\Sigma}}_s\boldsymbol{\Sigma}^{-1/2}$, without loss of generality we assume that
	$\boldsymbol{\Sigma}=\sigma^2\mathbf{I}_m$ \cite{Johnstone2017Biometrika}. Consequently, in what follows, we statistically characterize the maximum eigenvalue of  $ \widehat{\boldsymbol{\Theta}}$ for 
	\begin{align}
&p\widehat{\boldsymbol{\Sigma}}_n\sim\mathcal{CW}_m\left(p, \mathbf{I}_m\right)\\
	& n\widehat{\boldsymbol{\Sigma}}_s\sim\mathcal{SCW}_m\left(n, \mathbf{I}_m+\gamma \mathbf{s}\mathbf{s}^\dagger\right)
	\end{align}
	where $\gamma=\rho ||\mathbf{h}||^2/\sigma^2$ and $\mathbf{s}=\mathbf{h}/||\mathbf{h}||$ denotes a unit vector.   
	
	Let us denote the maximum eigenvalue of $\widehat{\boldsymbol{\Theta}}$ as $\hat{\lambda}_{\max}$. Then 
 the proposed test based on $\hat{\lambda}_{\max}$ detects a signal if 
 \begin{align}
     \hat{\lambda}_{\max}>\lambda_{\text{th}}
 \end{align}
 where $\lambda_{\text{th}}$ is the threshold corresponding to a desired false alarm rate\footnote{This is also known as the Type I error rate in the literature of statistics.} $\alpha\in(0,1)$ given by
 \begin{align}
\label{alarm}
\alpha=P_F(\lambda_{\text{th}})&=\Pr\left(\hat{\lambda}_{\max}>\lambda_{\text{th}}|\mathcal{H}_0\right). 
 \end{align}
 As usual, we choose $\lambda_{\text{th}}$ to maximize the 
 detection probability\footnote{Alternatively, this is called the power of the test.} given by \cite{ref:vantrees}
		\begin{align}
	 P_D(\gamma, \lambda_{\text{th}})&=\Pr\left(\hat{\lambda}_{\max}>\lambda_{\text{th}}|\mathcal{H}_1\right). \label{detection}
	\end{align}
Since the underlying probability density functions are continuous, the threshold evaluates to $\lambda_{\text{th}}=P_F^{-1}(\alpha)$. Now it is convenient to eliminate the $\lambda_{\text{th}}$ dependency to obtain a functional relationship between $P_D$ and $P_F$. 
In particular, the $(P_D, P_F) $ characterizes   the detector  and is referred to as the  ROC profile.
	
	The main technical challenge here is to statistically characterize the maximum eigenvalue of the {\it singular matrix} $\widehat{\boldsymbol{\Theta}}$, under the alternative $\mathcal{H}_1$, in terms of simple algebraic functions. Since the maximum eigenvalues of $\widehat{\boldsymbol{\Sigma}}_n^{-1/2}\widehat{\boldsymbol{\Sigma}}_s\widehat{\boldsymbol{\Sigma}}_n^{-1/2}$ and $\widehat{\boldsymbol{\Sigma}}_n^{-1}\widehat{\boldsymbol{\Sigma}}_s$ are the same, this amounts to determining the statistical characteristics of the maximum eigenvalue of $\widehat{\boldsymbol{\Sigma}}_n^{-1}\widehat{\boldsymbol{\Sigma}}_s$ which is also known as the {\it singular} $F$-matrix in the multivariate statistical literature \cite{shimizu2022expressing}, \cite{memuirhead2009aspects}. Although various statistical characteristics of the eigenvalues of {\it non-singular}-$F$ matrices are well documented in the literature (see e.g., \cite{Johnstone2008stat, jiang2013Bernoulli,Johnstone2017Biometrika,memuirhead2009aspects,meJames,Dumitriu2008siam,Dumitriu2012phys,dharmawansa2019roy,dubbs2014beta,kan2019densities} and references therein), a little has been documented related to the {\it singular}-$F$ matrices with a few exceptions. For instance, the joint eigenvalue density of {\it singular} $F$-matrices has been derived in \cite{shimizu2022expressing,srivastava2003singular,uhlig1994singular,diaz1997proof}. 
 In this respect, \cite{shimizu2022expressing} has expressed the c.d.f. of the maximum eigenvalue of singular $F$-matrix under both hypotheses in terms of heterogeneous hypergeometric function of two matrix arguments; the numerical complexity of which impede the utility of these expressions in other potential application domains. Moreover, the authors have provided an alternative expression involving hypergeometric function of one matrix argument for the same c.d.f. This latter form has a determinant representation involving a square matrix of size $m$ \cite{Khatri}. Also, when the argument matrix has a single spike, which is the case of our interest, this determinantal representation gives an indeterminate form. 
To circumvent this difficulty, Khatri \cite{Khatri} devised a technique based on the repeated application of l'Hospital's rule which also does not change the $m$ dependent matrix size. Since we have three main dimension-related parameters (i.e., $m,n$ and $p$) in our problem, sometimes it is convenient to have a determinant representation involving a matrix whose size depends on the relative difference between certain parameters rather then their individual magnitudes. For instance, determinant of a matrix whose size is $p-m$ would give more insights than a matrix of size $m$. From purely a computational point of view, the former structure reduces to a scalar when $p=m$, whereas the latter does not have that advantage. On the other hand, as we are well aware of, the difference $p-m$ is an implicit measure of  the quality of  $\widehat{\boldsymbol{\Sigma}}_n$ as an estimator of the unknown population noise covariance. In this respect, $p=m$ scenario dictates a performance lower bound on the leading eigenvalue as a test statistic. This representation also facilitates an asymptotic analysis of the ROC in the regime in which $m\to\infty$ such that $p/m\to 1$.

Having motivated with the above facts, in what follows, taking advantage of the contour integral representation of certain unitary integrals \cite{Mo2012,meWang}, we first derive a simple closed-form expression for the joint density of the non-zero eigenvalues of a complex correlated singular $F$-matrix with a single spiked associated covariance structure. Subsequently, capitalizing on this new joint density, we obtain an exact closed-form solution for the c.d.f. of the maximum eigenvalue in terms of easily computable functions by leveraging the powerful orthogonal polynomial approach due to Mehta \cite{me11}. Consequently this expression has been utilized to characterize the ROC of the proposed test.

	\section{C.D.F. of the Maximum Eigenvalue}

Here we develop some new fundamental results pertaining to the representation of the joint eigenvalue density of a complex correlated {\it singular} $F$-matrix and the c.d.f. of its leading eigenvalue. To this end, we require some preliminary results given below.
    
    
    \subsection{Preliminaries}

Let $\mathbf{A}\sim\mathcal{SCW}_m\left(n,\boldsymbol{\Sigma}\right)$ and $\mathbf{B}\sim\mathcal{CW}_m\left(p,\mathbf{I}_m\right)$ be two independent Wishart matrices with $p\geq m>n$. Then the matrix $\mathbf{A}$ is said to follow a {\it singular} Wishart matrix. As such, the density of $\mathbf{A}$ is defined on the space of $m\times m$ Hermitian positive {\it semi-definite} matrices of rank $n$ \cite{ratnarajah2005complex, Onatski2008Ann }. Now the matrix $\mathbf{F}=\mathbf{B}^{-1/2}\mathbf{A}\mathbf{B}^{-1/2}\in\mathbb{C}^{m\times m}$ follows a complex correlated {\it singular} $F$-distribution~\cite{shimizu2022expressing}. Therefore, $\mathbf{F}$ assumes the eigen-decomposition $\mathbf{F}=\mathbf{U}_1\boldsymbol{\Lambda}\mathbf{U}_1^\dagger$, where $\mathbf{U}_1\in\mathcal{V}_{n,m}$  and $\boldsymbol{\Lambda}=\text{diag}\left(\lambda_1, \lambda_2,\ldots,\lambda_n\right)$ denotes the non-zero eigenvalues of $\mathbf{F}$ ordered such that $0<\lambda_1<\lambda_2<\ldots<\lambda_n<\infty$.

\begin{thm}\label{F_Def}
The joint density of the ordered non-zero eigenvalues $0<\lambda_1<\lambda_2<\ldots<\lambda_n<\infty $ of the complex correlated {\it singular} matrix $\mathbf{F}$ is given by \cite{shimizu2022expressing}
\begin{align}
\label{eig_joint}
f(\lambda_1,\cdots,\lambda_n) =\frac{\mathcal{K}_1(m,n,p) }{\text{det}^n\left[\boldsymbol{\Sigma}\right]}
	\prod_{j=1}^n \lambda_j^{m-n}\Delta_n^2(\boldsymbol{\lambda})\int_{\mathcal{V}_{n,m}} \frac{
 \left(\mathbf{U}_1^\dagger {\rm d}\mathbf{U}_1\right)}{\det^{(n+p)}{\left[\mathbf{I}_m+\boldsymbol{\Sigma}^{-1}\mathbf{U}_1 \boldsymbol{\Lambda}\mathbf{U}_1^\dagger\right]}}
\end{align}
where $\left(\mathbf{U}_1^\dagger {\rm d}\mathbf{U}_1\right)$ denotes the exterior differential form representing the uniform measure on the complex Stiefel manifold~\cite{ratnarajah2005complex,Onatski2008Ann}, $\Delta_n(\boldsymbol{\lambda})=\prod_{1\leq i<j\leq n}\left(\lambda_j-\lambda_i\right)$ is the Vandermonde determinant, and 
\begin{align*}
\mathcal{K}_1(m,n,p)=\frac{\pi^{n(n-m-1)}\widetilde{\Gamma}_m(n+p)}{2^n\widetilde{\Gamma}_m(p)\widetilde{\Gamma}_n(n)}
\end{align*}
with the complex multivariate gamma function is written in terms of the classical gamma function $\Gamma(\cdot)$, for $M\in\mathbb{Z}^+$, as
\begin{align*}
\widetilde{\Gamma}_M(z)=\pi^{\frac{1}{2}M(M-1)} \prod_{j=1}^M \Gamma\left(z-j+1\right),\;\; \Re{\left\{z\right\}}>M-1
\end{align*}
\end{thm}
The following contour integral representation of the Harish-Chandra-Itzykson-Zuber integral \cite{harish1957differential,itzykson1980planar} for a rank deficient argument matrix due to \cite{Mo2012,meWang} is also instrumental in the sequel.\footnote{ A more generalized contour integral representation in this respect can be found in \cite{ref:peterJ}, \cite{ref:peter22}. However, the above form is adequate for our requirement.}
\begin{thm}
\label{thmcontour}
    Let $\mathbf{P}=\mathbf{p}\mathbf{p}^\dagger$ with $\mathbf{p}\in\mathbb{C}^{m\times 1}$ and $||\mathbf{p}||=1$. Let $\mathbf{T}\in\mathbb{C}^{m\times m}$ be Hermitian positive semi-definite with rank $n(<m)$. Then we have \cite{Mo2012,meWang}
    \begin{align}
    \label{eq cont integral}
       \int\limits_{\mathcal{U}_m} \text{etr}\left(\mathbf{P}\mathbf{U}\mathbf{TU}^\dagger\right) {\rm d}\mathbf{U}=\frac{\Gamma(m)}{2\pi \mathrm{i}} 
       \oint\limits_{\mathcal{C}}
        \frac{e^z}{z^{m-n}\displaystyle \prod_{j=1}^n \left(z-\tau_j\right)} {\rm d}z
    \end{align}
    where ${\rm d}\mathbf{U}$ is the invariant measure on the unitary group $\mathcal{U}_m$ normalized to make the total measure unity (i.e., $\int_{\mathcal{U}_m} {\rm d}\mathbf{U}=1$), $\tau_1,\tau_2,\ldots,\tau_n$ are the {\it non-zero} eigenvalues of $\mathbf{T}$, the contour $\mathcal{C}$ is large enough so that all $\tau_j$'s and $0$ are in its interior, and $\rm i=\sqrt{-1}$.
\end{thm}
Jacobi polynomials are intimately coupled with the Jacobi unitary ensemble. Therefore, one of the useful definitions of the Jacobi polynomial is given below.
\begin{defn}\label{defJac}
Jacobi polynomials can be written as follows \cite[eq. 8.962]{ref:gradshteyn}:

    \begin{equation}\label{jacobidef}
P_{n}^{(a,b)}(x) = \frac{(a+1)_n}{n!}\Hypergeometric{2}{1}{-n,n+a+b+1}{1+a}{\frac{1-x}{2}}
	\end{equation}
    where $a,b>-1$, ${}_2F_1 (\cdot,\cdot;\cdot;\cdot)$ denotes the Gauss hypergeometric function, and $(a)_k=a(a+1)\ldots(a+k-1)$ with $(a)_0=1$ denotes the Pochhammer symbol. Moreover, The Gauss hypergeometric function assumes the following expansion \cite{ref:gradshteyn}
    \begin{align}
    \label{Gauss}
        {}_2F_1(q,r;s;z)=\sum_{k=0}^\infty \frac{(q)_k (r)_k}{k! (s)_k} z^k,\;\;\; |z|<1
    \end{align}
    where $q,r,s\in\mathbb{C}$ with $s\notin \mathbb{Z}^-$.
    \end{defn}
    Consequently, the successive derivatives of the Jacobi polynomials can be written as
    \begin{equation}\label{jacobiDerivative}
	\frac{{\rm d}^{k}}{{\rm d}x^{k}}P_{n}^{(a,b)}(x) = 2^{-k}(n+a+b+1)_{k}P_{n-k}^{(a+k,b+k)}(x).
	\end{equation} 
    Finally, for a negative integer $-n$ with $n\in \mathbb{Z}^+$, we have \cite{me12}
    \begin{align}
    \label{apple}
    (-n)_k=\left\{\begin{array}{ll}
	\frac{(-1)^{k}n!}{(n-k)!} &  \text{if } 0\leq k\leq n\\
	0 & \text{if } k> n.
    \end{array}\right.
    \end{align}
\subsection{New Finite Dimensional  C.D.F.}
Having presented the above preliminary results, now we focus on deriving a new exact c.d.f. for the maximum eigenvalue of $\mathbf{F}$ when the covariance matrix $\boldsymbol{\Sigma}$ takes the so called rank-$1$ perturbation of the identity (i.e., single spiked) form. In this case, the covariance matrix can be decomposed as
\begin{align}
\label{spike_cov}
\boldsymbol{\Sigma}=\mathbf{I}_m+\eta \mathbf{ss}^\dagger=\mathbf{S}\hspace{0.4mm}\text{diag}\left(1+\eta, 1, 1,\ldots, 1\right)\mathbf{S}^\dagger 
\end{align}
from which we obtain
\begin{align}
\label{spike_invcov}
\boldsymbol{\Sigma}^{-1}=\left(\mathbf{I}_m+\eta \mathbf{ss}^\dagger\right)^{-1}= \mathbf{I}_m-\frac{\eta}{1+\eta} \mathbf{ss}^\dagger=\mathbf{I}_m-\mathbf{S}\boldsymbol{\Lambda}_\eta\mathbf{S}^\dagger
\end{align}
where $\mathbf{S}=\left(\mathbf{s}\; \mathbf{s}_2\; \ldots \mathbf{s}_m\right)\in\mathcal{U}_m$, $\boldsymbol{\Lambda}_\eta=\text{diag}\left(\eta/\eta+1,0,0,\cdots,0\right)$, and  $\eta\geq 0$. 
Following~\cite{shimizu2022expressing}, the matrix integral in (\ref{eig_joint}) can be expressed in terms of the so called {\it heterogeneous hypergeometric function of two matrix arguments} (see e.g., Theorem 2 therein). However, the utility of such functions are limited as they are not amenable to further analysis. To circumvent this difficulty, capitalizing on a contour integral approach due to \cite{Mo2012,meWang}, here we derive a new joint eigenvalue density which contains simple algebraic/transcendental functions. This new joint density further facilitates the use of powerful orthogonal polynomial techniques due to Mehta~\cite{me11} to derive the c.d.f. of the dominant eigenvalue.
The following corollary gives the new alternative expression for the joint density.

 \begin{cor}\label{joint_eig_pdf}
 Let $\mathbf{A}\sim\mathcal{SCW}_m(n,\mathbf{I}_m+\eta \mathbf{s}\mathbf{s}^\dagger)$ and $\mathbf{B}\sim\mathcal{CW}_m(p,\mathbf{I}_m)$ be independent Wishart matrices with $ p\geq m>n$ and $\eta> 0$. Then the joint density of the ordered non-zero eigenvalues $0\leq \lambda_1\leq \lambda_2\leq\cdots\leq \lambda_n<\infty$ of the {\it singular} matrix $\mathbf{F}=\mathbf{B}^{-1/2}\mathbf{A}\mathbf{B}^{-1/2}$ is given by
		\begin{align}
		\label{jpdf}
		f(\lambda_1,\cdots,\lambda_n)&=
		\frac{\mathcal{K}_2(m,n,p)}{\left(1+\eta\right)^n}
		\prod_{j=1}^n \frac{\lambda_j^{m-n}}{(1+\lambda_j)^{p+n}} \Delta_n^2(\boldsymbol{\lambda}) g\left(\frac{\lambda_1}{1+\lambda_1},\ldots, \frac{\lambda_n}{1+\lambda_n}\right)
		\end{align}
  where 
  \begin{align}
\label{eq_g}
    &g(x_1,\ldots,x_n)\nonumber\\
    & \qquad=\displaystyle \sum_{k=1}^n
    \frac{1}{\displaystyle \prod_{\substack{\ell=1\\ \ell\neq k}}^{n}\left(x_k-x_\ell\right)}\left[\frac{\Gamma(n+p-m+1)}{c_\eta^{m-1} x_k^{m-n}\left(1-c_\eta x_k\right)^{n+p-m+1}}-\sum_{j=0}^{m-n-1} \frac{\Gamma(p-j)}{\Gamma(m-n-j)c_\eta^{n+j}x_k^{j+1}}\right]
\end{align}
   with $c_\eta=\frac{\eta}{\eta+1}$ and $\mathcal{K}_2(m,n,p)=\frac{\pi^{n(n-1)}\Gamma(m)\widetilde{\Gamma}_m(n+p)}{\Gamma(n+p)\widetilde{\Gamma}_m(p)\widetilde{\Gamma}_n(n)\widetilde{\Gamma}_n(m)}$.
 \end{cor}
\begin{IEEEproof} 
See Appendix \ref{appjoint}.
\end{IEEEproof}
        \begin{rk}
		It is worth noting that the joint density corresponding to $\eta=0$ can easily  be obtained from (\ref{eig_joint}) as
  \begin{align}
  \label{pdf_eta0}
h(\lambda_1,\ldots,\lambda_n)=\frac{\pi^{n(n-1)}\widetilde{\Gamma}_m(n+p)}{\widetilde{\Gamma}_m(p)\widetilde{\Gamma}_n(n)\widetilde{\Gamma}_n(m) }\prod_{j=1}^n \frac{\lambda_j^{m-n}}{(1+\lambda_j)^{p+n}} \Delta_n^2(\boldsymbol{\lambda})
  \end{align}
  where we have used the fact $\int_{\mathcal{V}_{n,m}}\left(\mathbf{U}_1{\rm d}\mathbf{U}_1^\dagger\right)=\frac{2^n \pi^{mn}}{\widetilde{\Gamma}_n(m)}$. The above expression coincides with \cite[Corollary 1]{shimizu2022expressing}.
	\end{rk}

   We may use the new join density given in Corollary \ref{joint_eig_pdf} to obtain the  c.d.f. of the maximum eigenvalue of singular $F$-matrix which is one of the main contributions of this paper.  To this end, by definition, the c.d.f. of the maximum eigenvalue can be written as
   \begin{align}
   \label{cdfdef}
	\Pr(\lambda_{\max}\leq x)=\idotsint\limits_{0\leq \lambda_{1}\leq \lambda_{2} \leq \cdots\leq \lambda_{n}\leq x} 
	 f(\lambda_1,\lambda_2,\cdots,\lambda_n) {\rm d}\lambda_1 {\rm d}\lambda_2\cdots{\rm d}\lambda_n.
	\end{align}
    The above multi-dimensional integral can be evaluated as shown in Appendix \ref{appcdf} to obtain the c.d.f. of the maximum eigenvalue which is given by the following theorem.
	\begin{thm}\label{newcdf}
		Let $\mathbf{A}\sim\mathcal{SCW}_m(n,\mathbf{I}_m+\eta \mathbf{ss}^\dagger)$ and $\mathbf{B}\sim\mathcal{CW}_m(p,\mathbf{I}_m)$ be independent with $p\geq m>n$ and $\eta> 0$. Then the c.d.f. of the maximum eigenvalue $ \lambda_{\max} $ of the {\it singular} matrix $\mathbf{F}=\mathbf{B}^{-1/2}\mathbf{A}\mathbf{B}^{-1/2}$ is given by
		\begin{align*}
			F^{(\alpha)}_{\lambda_{\max}}(x;\eta)
			&=\dfrac{\mathcal{K}_\alpha(m,n)}{(1+\eta)^{n}}
   G^{(\alpha)}_\eta\left(\frac{x}{1+x}\right)
		\end{align*}
		where 
  \begin{align}
\label{eq G defn}
   G^{(\alpha)}_\eta(x)&=\frac{(n+\alpha)!x^{n(\alpha+m)-m+1}}{c_\eta^{m-1}\left(1-c_\eta x\right)^{n+\alpha+1}}\det\left[\Omega^{(\alpha)}_{i}(x,\eta)\hspace{3mm} \Psi_{i,j}(x)\right]_{\substack{i=1,2,...,\alpha+1\\j=2,3,...,\alpha+1}}\nonumber\\
   &\hspace{4cm}
   +\frac{(-1)^n}{c_\eta^n}x^{n(m+\alpha-1)}
   \det\left[(-1)^{i-1}\Phi^{(\alpha)}_{i}(x,\eta)\hspace{3mm} \Psi_{i,j}(x)\right]_{\substack{i=1,2,...,\alpha+1\\j=2,3,...,\alpha+1}},
\end{align}
		\begin{align*}
  &\Psi_{i,j}(x)= (m+i-1)_{j-2}P_{n+i-j}^{(j-2,m-n+j-2)}\left(\frac{2}{x}-1\right),\\
  & \Phi^{(\alpha)}_i(x,\eta)=\sum_{k=0}^{m-n-1} \frac{(m+\alpha-k-1)!(n+k+i-2)!}{k!(m+i-k-2)!c_\eta^k x^k},\\
		&\Omega^{(\alpha)}_{i}(x,\eta)=\frac{(n+i-2)!}{(m+i-2)!}\sum_{k=0}^{n+i-2}
  \frac{(-1)^k(m+i+k-2)!}{(n+i-2-k)!k!(k+1)!}\\
  & \hspace{5cm}\times 
  {}_2F_1\left(n+\alpha+1,k+1;k+2;\frac{-x\eta}{1+\eta(1-x)}\right),
		\end{align*}
$\alpha=p-m$, and $\mathcal{K}_\alpha(m,n)=\displaystyle \prod_{j=1}^{\alpha}\frac{(m+n+j-2)!}{(n-1)!(m+n+2j-2)!}$.
	\end{thm}
	\begin{IEEEproof}
	See Appendix \ref{appcdf}.
	\end{IEEEproof}
It is noteworthy that the above theorem holds under the strict condition $\eta>0$. Therefore, the case corresponding to $\eta=0$ is given by the following theorem.
\begin{thm} \label{thmnull}
    Let $\mathbf{A}\sim\mathcal{SCW}_m(n,\mathbf{I}_m)$ and $\mathbf{B}\sim\mathcal{CW}_m(p,\mathbf{I}_m)$ be independent with $p\geq m>n$. Then the c.d.f. of the maximum eigenvalue $ \lambda_{\max} $ of the {\it singular} matrix $\mathbf{F}=\mathbf{B}^{-1/2}\mathbf{A}\mathbf{B}^{-1/2}$ is given by
   \begin{align}\label{cdfeta0}
			F^{(\alpha)}_{\lambda_{\max}}(x;0)
			&=\prod_{k=1}^\alpha \frac{(m+n+k-1)!}{(m+n+2k-2)!}\left(\dfrac{x}{1+x}\right)^{n(m+\alpha)}\det\left[\Psi_{i+1,j+1}\left(\frac{x}{1+x}\right)\right]_{i,j=1,2,...,\alpha}.
		\end{align}
\end{thm}
\begin{IEEEproof}
    See Appendix \ref{appcdfeta0}.
\end{IEEEproof}

The computational complexity of the above new c.d.f.s depends on the size of the determinant which is $\alpha=p-m$. Clearly, when the relative difference between $p$ and $m$ is small, irrespective of their individual magnitudes,  the c.d.f.s can be computed very efficiently.  This distinctive advantage is due to the orthogonal polynomial approach that we have employed in statistically characterizing the maximum eigenvalue. To further highlight this fact, in the following corollary, we present the c.d.f.s corresponding to the special configuration $\alpha=0$.
\begin{cor}\label{coralpha0}
		The exact c.d.f.s of the maximum eigenvalues of $\mathbf{B}^{-1/2}\mathbf{A}\mathbf{B}^{-1/2}$ corresponding to $\alpha= 0$ (i.e., $m=p$) for $\eta>0$ and $\eta=0$ are given, respectively, by
		\begin{align}
			F^{(0)}_{\lambda_{\max}}(x;\eta)=&\frac{n!(1+\eta)^{m} x^{m(n-1)+1}}{(m-1)! \eta^{m-1}(1+x)^{mn-m-n}\left(1+\eta+ x \right)^{n+1}}\nonumber\\
   & \qquad \times \sum_{k=0}^{n-1}
   \frac{(-1)^k (m+k-1)!}{k! (k+1)! (n-k-1)!}\hspace{1mm}{}_2F_1\left(n+1,k+1;k+2;-\frac{\eta x}{1+\eta+x}\right)\nonumber\\
   & \qquad \qquad \qquad \qquad \qquad + \frac{(-1)^n}{(n-1)! \eta^{n} } \sum_{k=0}^{m-n-1} \frac{(n+k-1)!x^{n(m-1)-k}}{k! c_\eta^k(1+x)^{n(m-1)-k}},\label{eq_cdf0}\\
   F^{(0)}_{\lambda_{\max}}(x;0)=&\dfrac{x^{mn}}{(1+x)^{mn}}.
		\end{align}
	\end{cor}  
\begin{IEEEproof}
    The proof follows by noting that the corresponding determinants degenerate to scalars for $\alpha=0$.
\end{IEEEproof}
  It is noteworthy that the hypergeometric function in the above expression degenerates to a finite series as shown below
  \begin{align}
  \label{eq_hyposimp}
      &{}_2F_1\left(n+1,k+1;k+2;-\frac{\eta x}{1+\eta+x}\right)\nonumber\\
      &\hspace{5cm}=\frac{(1+\eta+x)^n}{(1+\eta)^n(1+x)^n} {}_2F_1\left(-(n-k-1),1;k+2;-\frac{\eta x}{1+\eta+x}\right)\nonumber\\
      &\hspace{5cm}=\frac{(1+\eta+x)^n}{(1+\eta)^n(1+x)^n} \sum_{\ell=0}^{n-1-k} (-1)^\ell
      \frac{(-(n-k-1))_\ell \hspace{1mm} \eta^\ell x^\ell}{(k+2)_\ell (1+\eta+x)^\ell \ell!} 
  \end{align}
  where the first equality follows from the hypergeometric transformation ${}_2F_1(a,b;c;z)=(1-z)^{c-a-b} {}_2F_1(c-a,c-b;c;z)$ \cite{ref:gradshteyn}  and the last equality is due to (\ref{Gauss}) and (\ref{apple}). To further highlight the effect of sample deficiency on the analytical complexity of the c.d.f. and for comparison, here we present the corresponding c.d.f. in the sample non-deficient scenario (i.e., $m\geq n$) as \cite[eq. 21] {chamain2020eigenvalue}
  \begin{align}
      F^{(0)}_{\lambda_{\max}}(x;\eta)
			=\dfrac{\left(\dfrac{x}{1+x}\right)^{mn}}{\left(1+\dfrac{\eta }{1+x}\right)^{n}},\;\;\;\; m\geq n,\; \eta\geq 0.
  \end{align}
  
\begin{figure}[t!]
	\centering
	\subfloat[CDF for different values of $n$ with $m=10$ and $p=15$. 
 ]{
		\label{fig_cdfnull_diff_n}
		\includegraphics[width=0.45\textwidth]{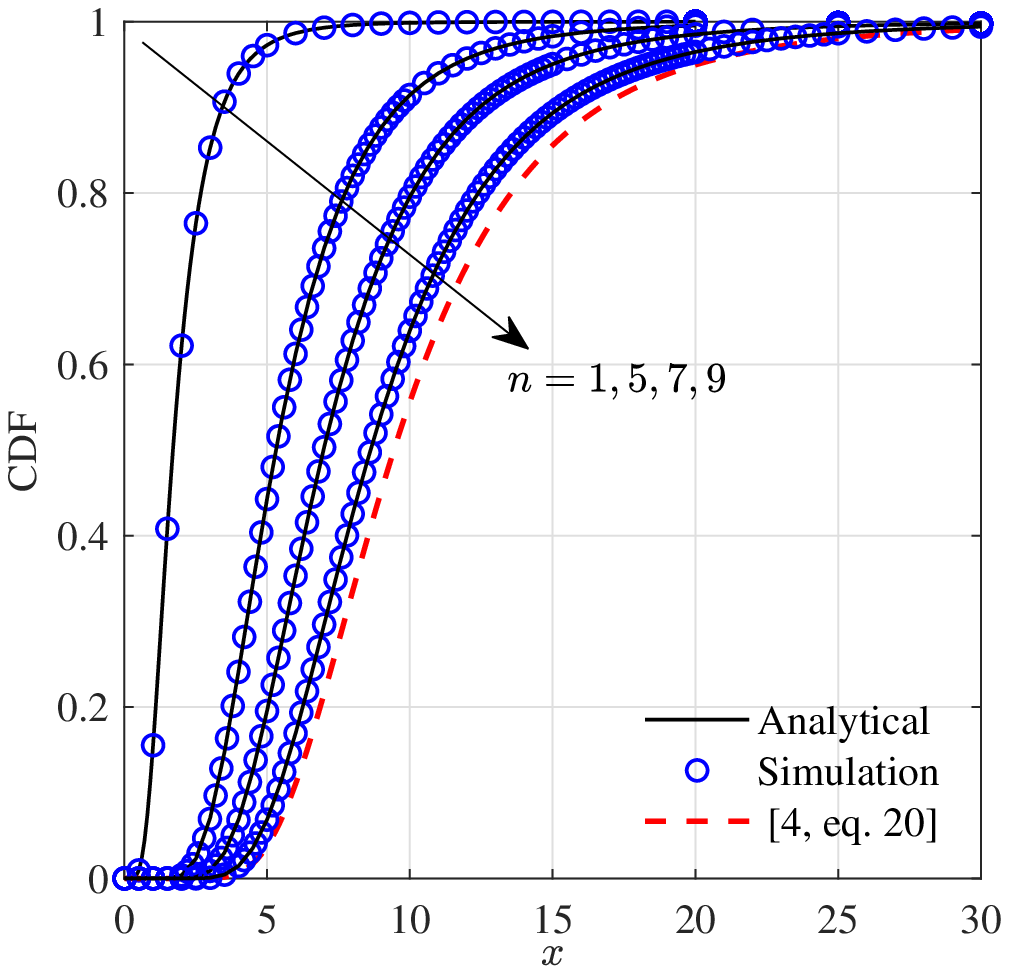}} 
	\subfloat[CDF for different values of $p$ with $n=5$ and $m=8$. 
 ]{
		\label{fig_cdfnull_diff_p}
		\includegraphics[width=0.45\textwidth]{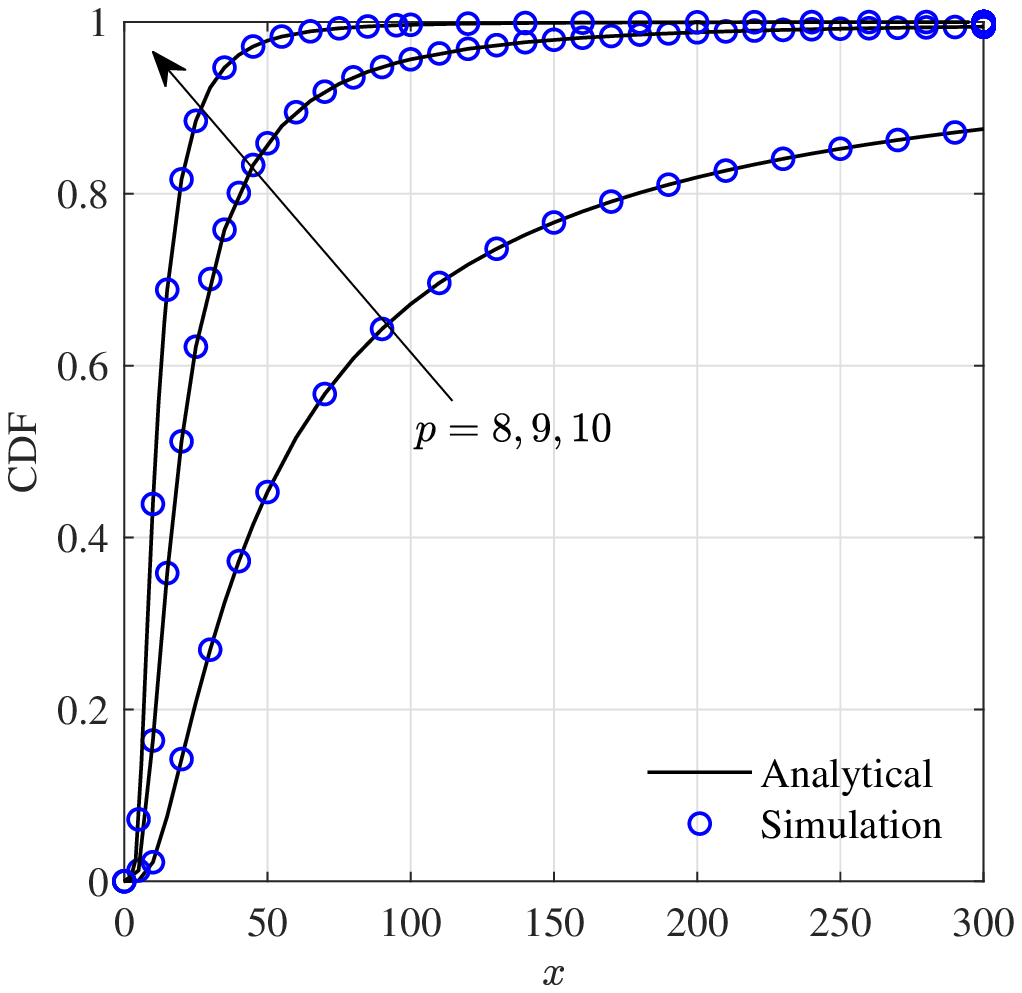}} 
	\caption{Comparison between the theoretical c.d.f. given in Theorem \ref{cdf_null} and simulated data points for various system configurations with $\eta=0$. The red dashed curve \cite[eq. 20]{chamain2020eigenvalue} depicts the case corresponding to the configuration $m=n$.
 }
	\label{cdf_null}
\end{figure}

\begin{figure}[t!]
	\centering
	\subfloat[CDF for different values of $n$. 
 ]{
		\label{fig_cdf_diff_n}
		\includegraphics[width=0.45\textwidth]{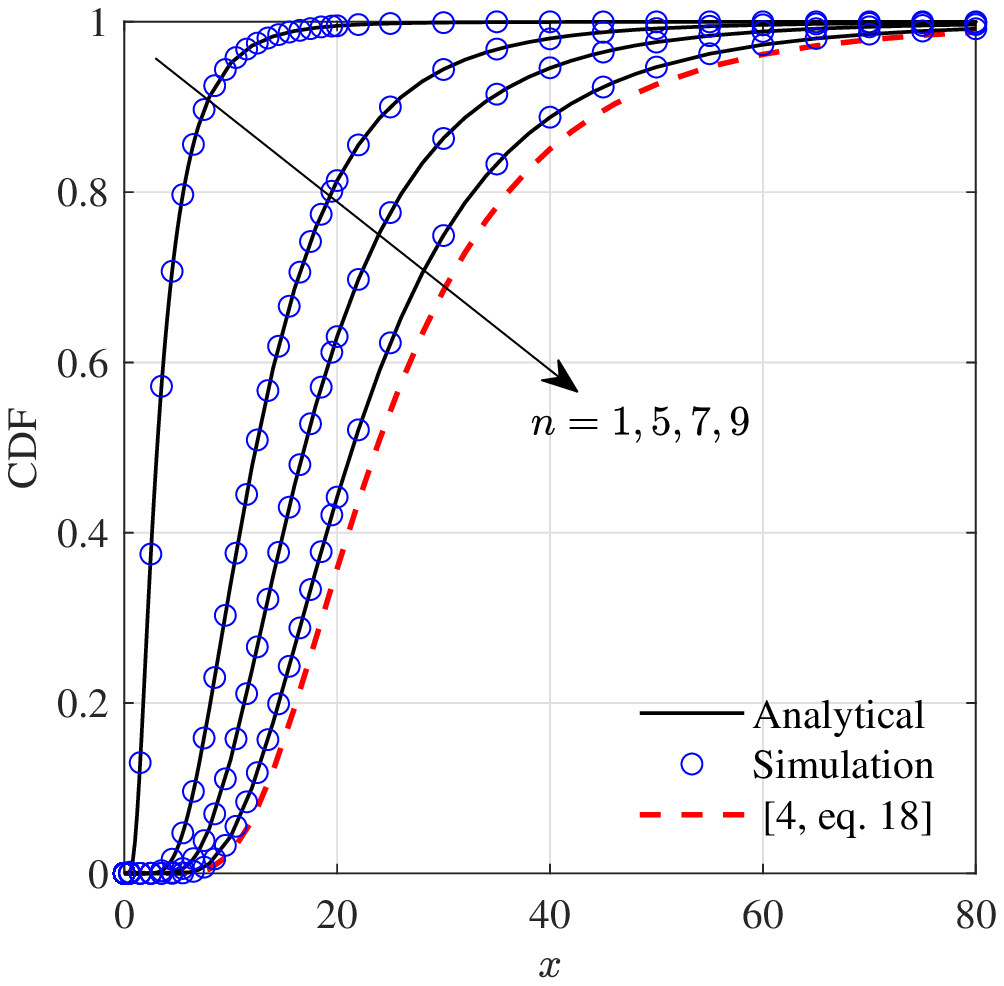}} 
	\subfloat[CDF for different values of $p$ with $n=5$ and $m=8$. 
 ]{
		\label{fig_cdf_diff_p}
		\includegraphics[width=0.45\textwidth]{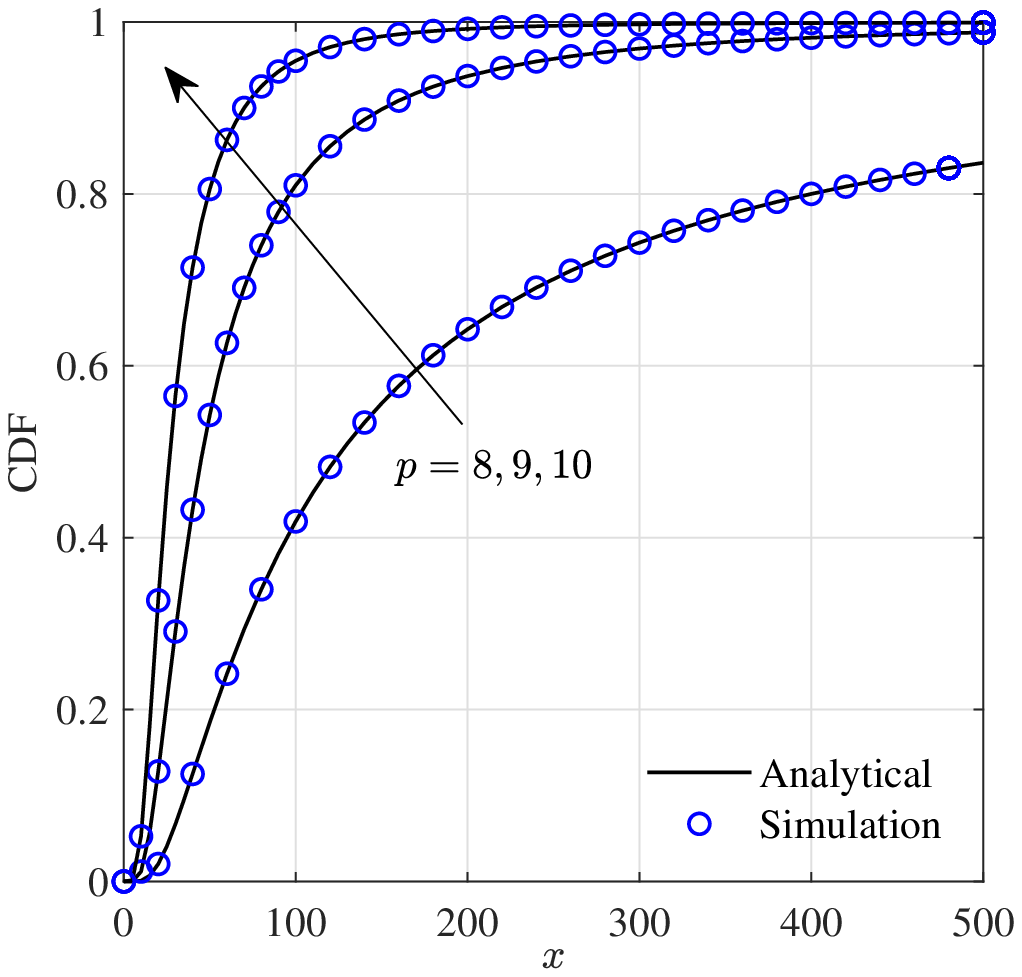}} \vspace{-2mm}
	\caption{Comparison between the theoretical c.d.f. in Theorem \ref{newcdf} with simulated values for various system configurations with $\eta=10\;$dB. The results are shown for $m=10$ and $p=15$. The red dashed curve \cite[eq. 18]{chamain2020eigenvalue} in Fig. 2a depicts the case corresponding to the configuration $m=n$.
 }\vspace{-2mm}
	\label{cdf}
\end{figure}

\begin{figure}[htp]
		\centering
	\includegraphics[width=0.45\textwidth]{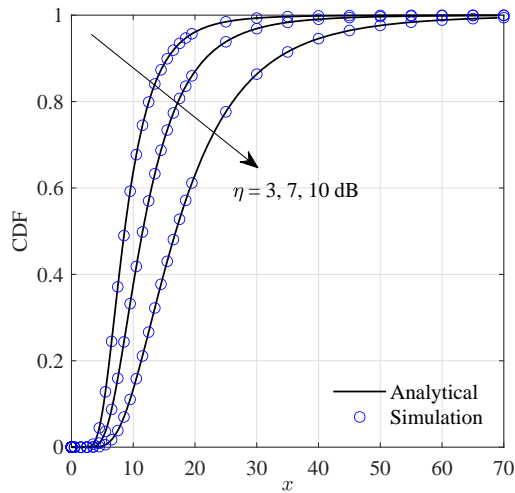}
		\caption{Comparison between the theoretical c.d.f. in Theorem \ref{cdf_null} and simulated data points for various system configurations with $\eta>0$. The results are shown for $m=10,\; n=7$, and $p=15$.
 }
		\label{fig_cdf_diff_eta}
	\end{figure}

Figure \ref{cdf_null} compares the analytical c.d.f. expression given by Theorem \ref{thmnull} with simulated data points for various system configurations under the condition that $\eta=0$. In particular, 
the effect of  $n$ on the c.d.f. for $m=10$ and $p=15$ is shown in Fig. \ref{fig_cdfnull_diff_n}. The red dashed curve, which is derived in \cite{chamain2020eigenvalue}, corresponding to $m=n$ case depicts the boundary between the full rank and rank deficient $\mathbf{A}$. On the other hand, the effect of $p$ is shown in  Fig.\ref{fig_cdfnull_diff_p}.  The 
curve corresponding to $p=m=8$ shows the boundary beyond which the matrix $\mathbf{B}$ assumes full rank. Figure 
\ref{cdf} depicts similar  dynamics corresponding to the scenario $\eta=10\;$dB. Finally, the effect of $\eta$--the strength of the rank-one spike--is depicted in Fig. \ref{fig_cdf_diff_eta}.

 Having armed with the above statistical characteristics of the maximum eigenvalue of complex correlated {\it singular} $\mathbf{B}^{-1/2}\mathbf{A}\mathbf{B}^{-1/2}$, in what follows, we focus on the ROC of the maximum eigenvalue based detector.

    \section{ROC of the Largest Generalized Eigenvalue}

    Let us now analyze the behavior of detection and false alarm probabilities associated with the maximum eigenvalue based test. To this end, by exploiting the relationship between the {\it non-zero} eigenvalues of $\widehat{\boldsymbol{\Theta}}$ and $\mathbf{F}$ given by by $\hat{\lambda}_j=(p/n)\lambda_j$, for $j=1,2,\ldots,n$, and keeping in mind the analogy between $\gamma$ and $\eta$, we may express the c.d.f. of the maximum eigenvalue corresponding to $\widehat{\boldsymbol{\Theta}}$ as
    $F_{\lambda_\max}^{(\alpha)}(\kappa x;\gamma)$, where $\kappa=n/p$.

    Now in light of Theorems \ref{newcdf} and \ref{thmnull}  along with (\ref{alarm}), (\ref{detection}), the detection and false alarm probabilities can be written, respectively, as
    \begin{align}
    P_D(\gamma, \lambda_{\text{th}})&=1-F_{\lambda_\max}^{(\alpha)}(\kappa\lambda_{\text{th}};\gamma)\\
    P_F(\lambda_{\text{th}})&=1-F_{\lambda_\max}^{(\alpha)}(\kappa\lambda_{\text{th}};0). 
    \end{align}
In general, obtaining an explicit functional relationship between $P_D$ and $P_F$ (i.e., the ROC profile) for an arbitrary system configuration is an arduous task. Nevertheless, in the important case of $\alpha=0$, such an explicit relationship is possible as shown in the following corollary.
    \begin{cor}
    \label{corasybalanced}
    In the important case of $\alpha=0$ (i.e., $p=m$),  the quantities $P_D$ and $P_F$ are functionally related as
    \begin{align}
    \label{rocbalanced}
    P_D=1-F_{\lambda_\max}^{(0)}\left[\frac{\left(1-P_F\right)^{1/nm}}
    {1-\left(1-P_F\right)^{1/nm}};\gamma\right].
    \end{align}
    \end{cor}
Since the configuration $p=m$ (i.e., when the number of noise only samples equals the system dimensionality or $\alpha=0$) barely guarantees the positive definiteness of the sample estimate of the noise-only covariance matrix\cite{memuirhead2009aspects}, this represents the worst possible ROC profile. Moreover, It turns out that, when $n=1$ (i.e., only one possible signal bearing sample is available), after some algebraic manipulations, (\ref{rocbalanced}) simplifies to
\begin{align}
     P_D=1-\frac{1-P_F}
    {1+\gamma-\gamma\left(1-P_F\right)^{1/m}},\;\;\; m\geq 2.
\end{align}
The above expression shows that as $m$ increases the power of the test decreases; eventually, it achieves the limit $P_D=P_F$ as $m\to\infty$. To further investigate the relationship between the power and $m$ for a single observation, we expand the power as
\begin{align}
    P_D=P_F-\left(1-P_F\right)\ln\left(1-P_F\right)\frac{\gamma}{m}+o\left(\frac{\gamma}{m}\right)
\end{align}
which clearly demonstrates how the power decays with $m$. Interestingly, the above expansion reveals that if $\gamma=O(m)$ then it is plausible that the power of the test converges to an asymptotic ROC profile other than $P_D=P_F$ as $m\to\infty$. It turns out that this insight, though obtained by analyzing the ROC profile corresponding to $n=1$, can be generalized to an arbitrary value of $n (<m)$ for $\alpha=0$ as given by the following corollary.
\begin{cor}\label{corasy}
    As $m,p,\gamma\to\infty$ such that $m/p\to 1$ and $\gamma/m\to c\in[0,\infty)$, the ROC admits the following asymptotic limit
    \begin{align}
    \label{cor asym limit}
        P_D^{\rm asy}=1-\frac{1-P_F}
    {\left[1-\displaystyle \frac{c}{n}\ln \left(1-P_F\right)\right]^n}.
    \end{align}
\end{cor}
\begin{IEEEproof}
    See Appendix \ref{app asym}.
\end{IEEEproof}
As a sanity check, when $c=0$ (i.e.,  $\gamma$ does not scale with $m$) the above asymptotic ROC degenerates to $P_D=P_F$, thereby confirming that the maximum eigenvalue has no detection power in this particular asymptotic regime. Similar conclusions have been made about the asymptotic detection power of the maximum eigenvalue of $F$-matrices in \cite{Nadakuditi2010jsacsp}, \cite{Johnstone2020stat}. The underpinning main technical argument for this loss of power is that, in this regime, the maximum eigenvalue lies below the so called phase-transition threshold and therefore, under the both hypotheses the  maximum eigenvalue (viz., properly centered and scaled maximum eigenvalue) converges to the same distribution--Tracy-Widom distribution \cite{Nadakuditi2010jsacsp}. In a sharp contrast,  Corollary \ref{corasy} shows that, even in a severely sample deficient scenario, the maximum eigenvalue retains its detection power asymptotically, given SNR is proportional to the system dimension. Moreover, noting that $\left[1-\displaystyle \frac{c}{n}\ln \left(1-P_F\right)\right]^n$ is an increasing function of $n$, we can  bound the asymptotic ROC as
\begin{align}
\label{eq upper asy}
    P_D^{\rm asy}\leq  1-\left(1-P_F\right)^{c+1},\;\;  c\geq 0
\end{align}
where the equality holds for $c=0$. Although the condition that the SNR being proportional to the system dimension seems to be too stringent, it is satisfied asymptotically by the most commonly used statistical fading channel model in the literature. To be specific,  when $\mathbf{h}\sim\mathcal{CN}\left(\mathbf{0}, \mathbf{I}_m\right)$ (i.e., Rayleigh channel), in view of the strong law of large numbers we get 
\begin{align}
    \lim_{m\to\infty}\frac{||\mathbf{h}||^2}{m}\to 1 \;\;\; \text{almost surely},
\end{align}
which in turn reveals how the SNR given by $\rho ||\mathbf{h}||^2/\sigma^2$ scales with $m$ asymptotically.
Therefore, the above asymptotic results can be of paramount importance in various modern wireless applications.

\begin{figure}[t!]
	\centering
	\subfloat[ROC for different values of $n$ with $\gamma=10,20$ {dB}. 
 ]{
		\label{fig_ROC_diff_n}
		\includegraphics[width=0.45\textwidth]{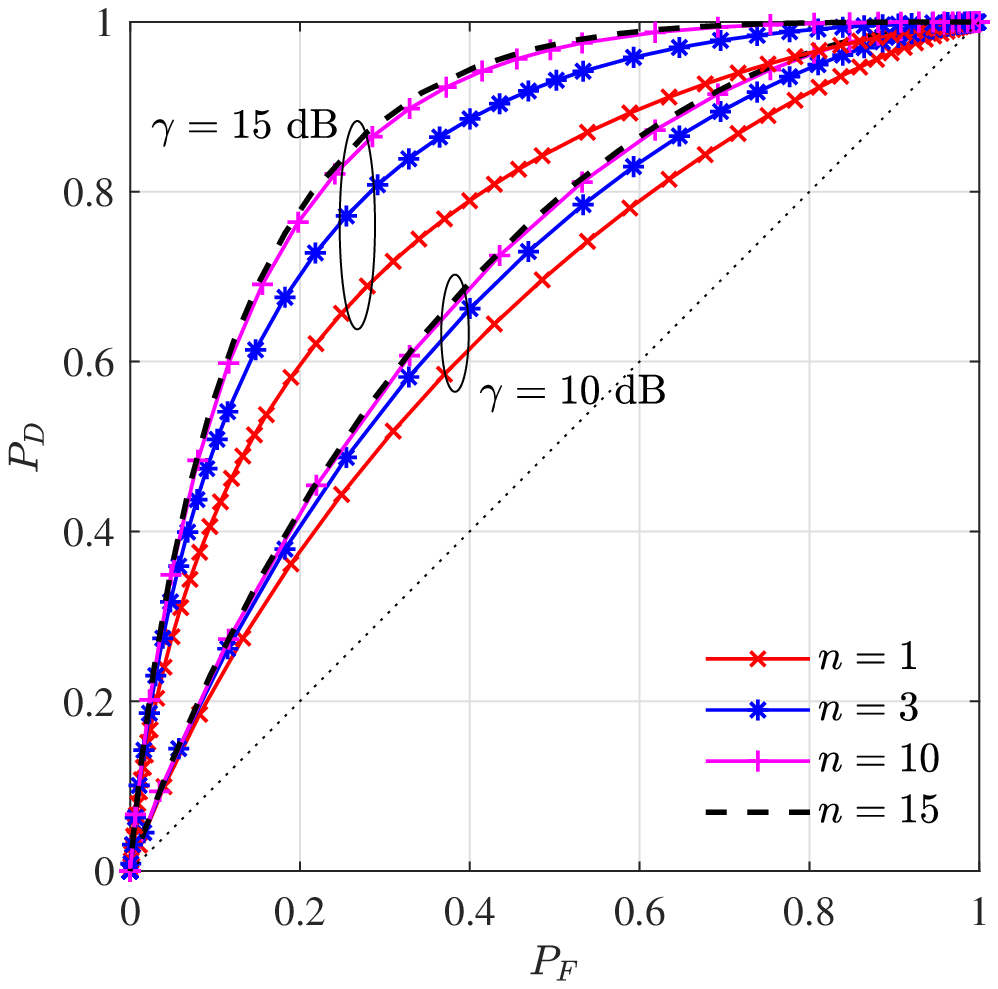}} 
	\subfloat[ROC for different values of $\gamma$ with $n=10$. 
 ]{
		\label{fig_ROC_diff_snr}
		\includegraphics[width=0.45\textwidth]{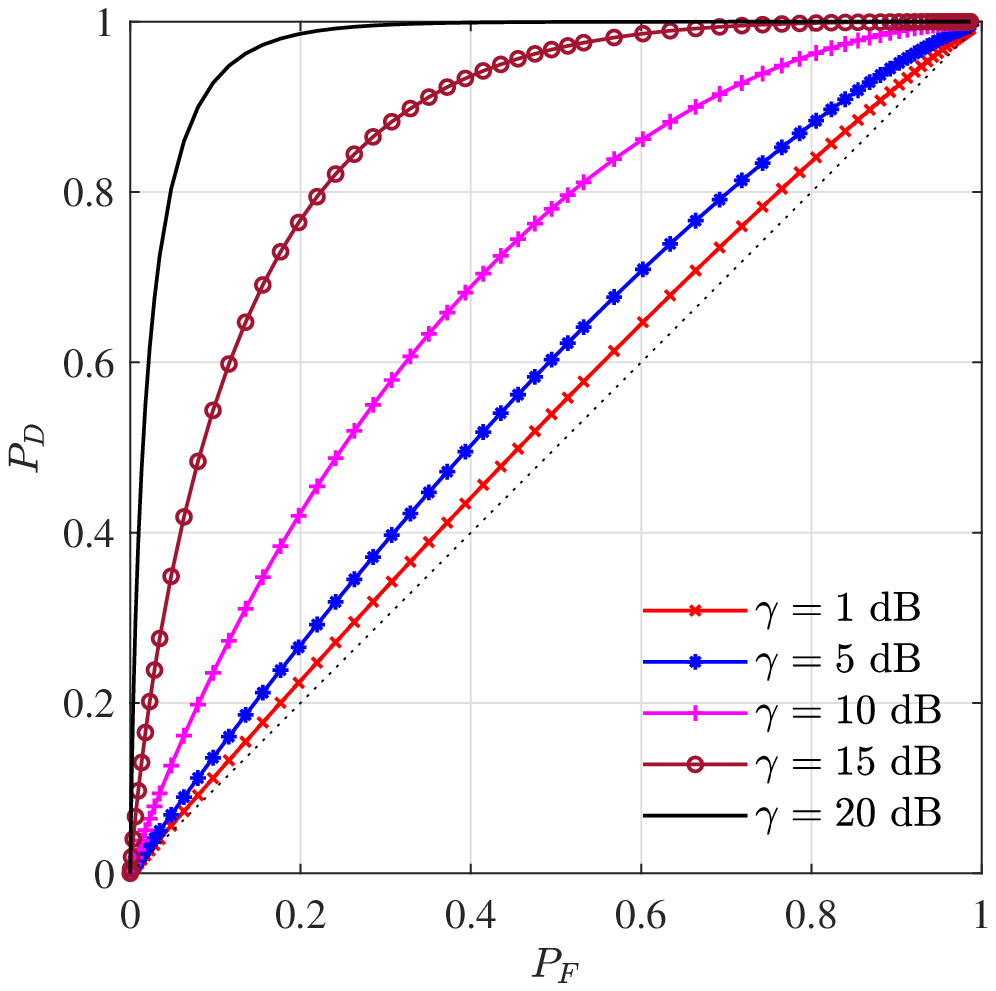}} 
	\caption{ The effect of $n$ and $\gamma$ on the ROC profiles. The results are shown for $m=15$ and $p=16$. The black dashed curves in Fig. 4a are generated with \cite[eqs. 24,25]{chamain2020eigenvalue}.
 }
	\label{fig_roc_diff_ng}
\end{figure}

The ROC dynamics that have been analytically quantified in the preceding discussion are numerically verified here. To be specific, 
Fig. \ref{fig_ROC_diff_n} shows the effect of plausible signal-plus-noise sample size $n$ on the power for various SNR values. The black dashed lines therein, which correspond to the case $m=n$, have been drawn based on \cite[eqs. 24,25]{chamain2020eigenvalue}.
Clearly, these black dashed lines mark the boundaries at which the sample deficient and non-deficient regions are separated.
As can be see from the graph, the disparity between $m$ and $n$ degrades the ROC profile uniformly for all SNR values. The effect of SNR on the power is  depicted in Fig. \ref{fig_roc_diff_ng}. 

\begin{figure}[t!]
	\centering
	\subfloat[ROC for different values of $m$ with  $p = 20$. 
 ]{
		\label{fig_ROC_diff_m}
		\includegraphics[width=0.45\textwidth]{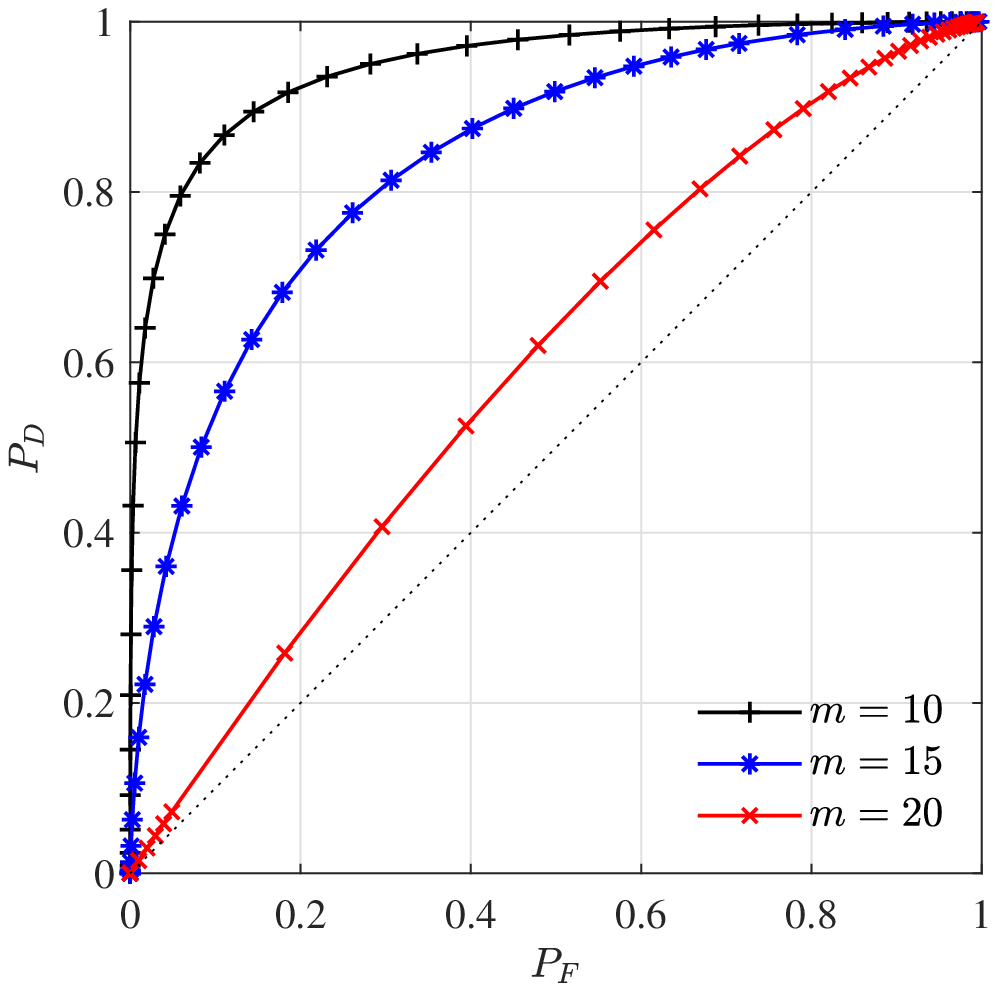}} 
	\subfloat[ROC for different $m=p$ configurations. 
 ]{
		\label{fig_ROC_diff_mp}
		\includegraphics[width=0.45\textwidth]{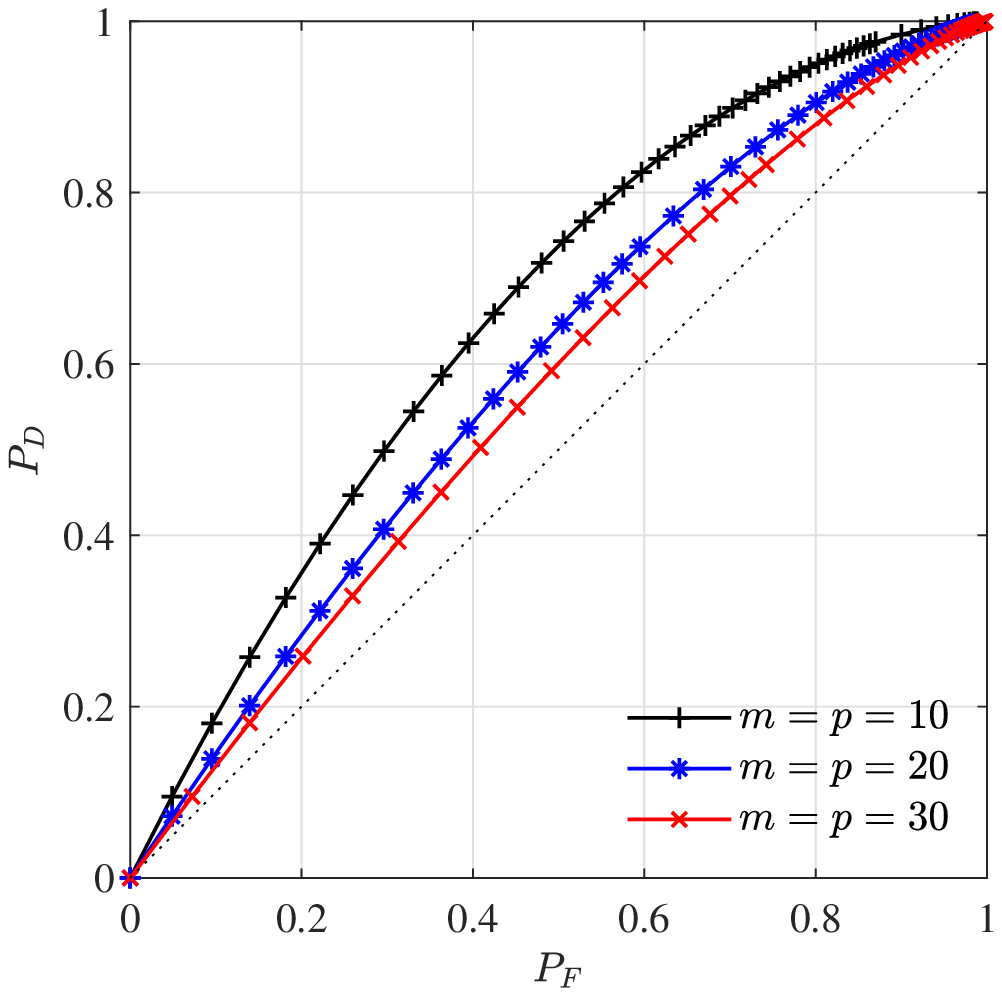}} 
	\caption{The effect of the system dimensionality $m$ on the ROC profiles. The results are shown for $n=5$ and $\gamma=10$ {dB}.
 }
	\label{fig_roc_mp}
\end{figure}

\begin{figure}[t!]
	\centering
	\subfloat[ROC for various $m=n$ configurations. 
 ]{
		\label{fig_ROC_diff_asy}
		\includegraphics[width=0.45\textwidth]{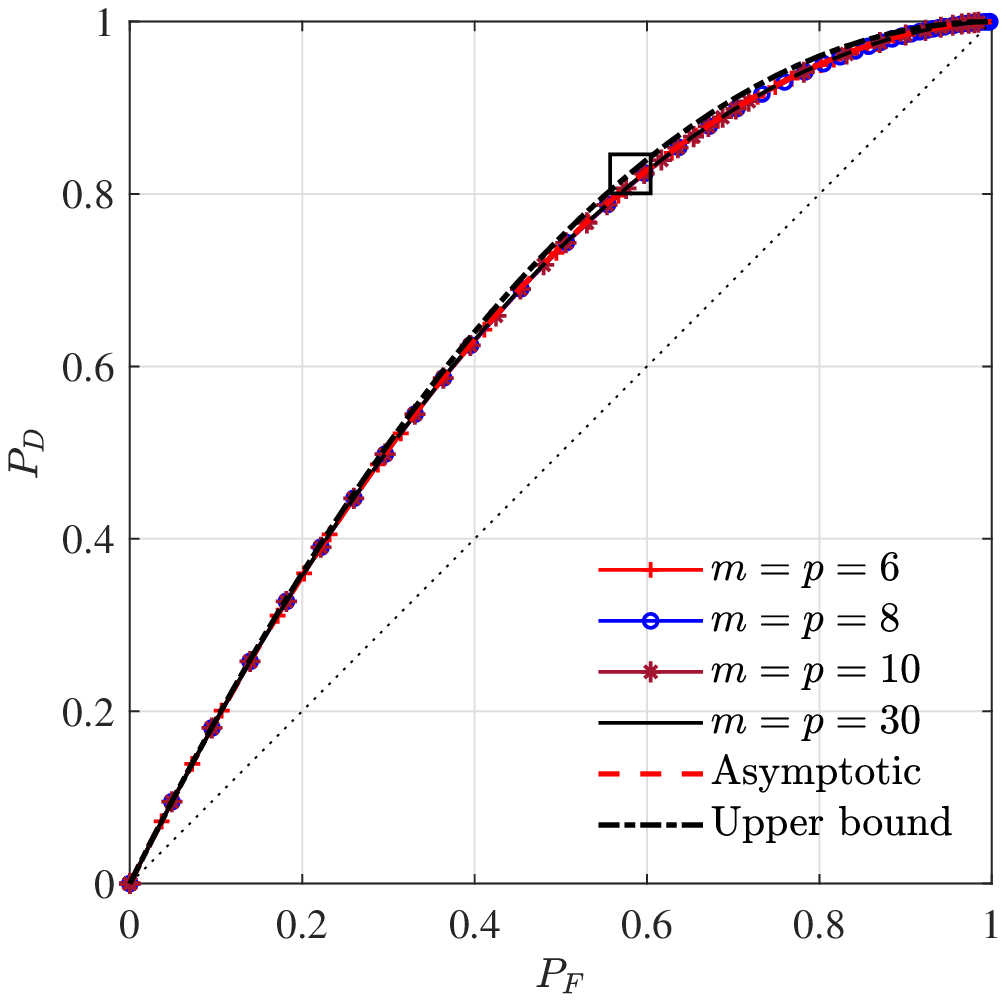}} 
	\subfloat[Magnified view of the small square on the left. 
 ]{
		\label{fig_ROC_diff_asy_zoom}
		\includegraphics[width=0.45\textwidth]{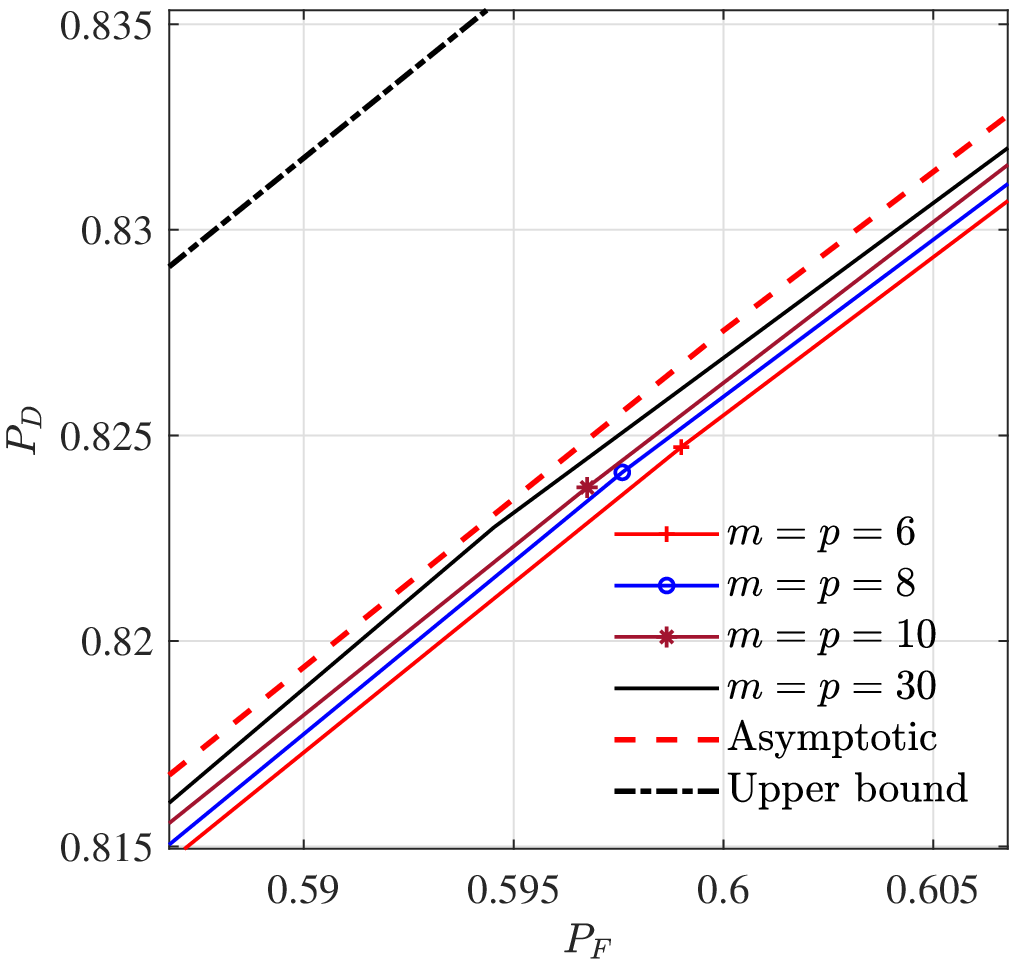}} 
	\caption{Comparison between the asymptotic ROC and its finite dimensional counterparts. Results are shown for $\gamma=m$ and $n=5$. Asymptotic ROC profile, as $m,p\to\infty$ such that $m/p\to 1$, is shown by the red dashed line, whereas the corresponding upper bound is depicted by the black dotdased line. 
 }
	\label{fig_roc_asy}
\end{figure}

Let us now numerically investigate the effect of system dimensionality $m$ on the ROC profile. To this end, we plot ROC profiles for different values of $m$ in Fig. \ref{fig_roc_mp}. Since $p$ is held fixed, the ROC profile degrades with the increasing $m$ as can be seen from Fig. \ref{fig_ROC_diff_m}. In this respect, the worst possible ROC profile is achieved  when $m=n$ as we have expected. This observation stems from the fact that the shrinking relative disparity between $p$ and $m$ tend to reduce the quality of the estimated noise covariance matrix. To further investigate the effect of the system dimensionality $m$ on the worst possible ROC profile, in Fig. \ref{fig_ROC_diff_mp}, we plot the the ROC profiles corresponding to $m=p$ scenario for various values of $m$. As we have expected, the increasing $m$ tend to degrade the worst profiles towards the chance curve $P_D=P_F$. This observation, although made based on the finite values of $m,p$, further verifies the fact that as $m,p\to\infty$ such that $m/p\to 1$, the leading generalized eigenvalue does not have the detection power. However, as Corollary \ref{corasy} states, when $\gamma=O(m)$, the leading eigenvalues retains its detection power in this particular asymptotic regime. These dynamics are depicted in Fig. \ref{fig_roc_asy}. For instance, as shown in Fig. \ref{fig_ROC_diff_asy}, although proved for asymptotically large values of $m,p$, the limiting ROC profile given in Corollary \ref{corasy} serves as a very good approximation to the ROC profile corresponding to $m,p$ values as little as $6$ when $\gamma=O(m)$. More interestingly, Fig. \ref{fig_ROC_diff_asy} illustrates that the upper bound (\ref{eq upper asy}) also serves as a good approximation to the finite $m=p$ configurations when $\gamma=O(m)$. Since the differences between the asymptotic  ROC profiles and that of profiles corresponding to their finite counterparts are indistinguishable in Fig. \ref{fig_ROC_diff_asy}. To further highlight this fact, in Fig. \ref{fig_ROC_diff_asy_zoom}, we present a magnified section of a small square marked in Fig. \ref{fig_ROC_diff_asy}. The tightness of the derived benchmark asymptotic ROC profiles is clearly visible in this figure, thereby verifying the accuracy of our formulation for small configurations as well.

\section{Conclusion}
This paper investigates the detection problem in colored noise using the largest generalized eigenvalue of whitened signal-plus-noise sample covariance matrix (a.k.a. $F$-matrix) as the test statistic. This $F$-matrix is endowed with a rank-one spiked underlying covariance structure due to the cumulative effects of whitening and the nature of the detection problem. Our specific focus is on the sample deficient regime in which the number of signal-plus-noise observations is strictly less than the system dimension (i.e., $n<m$). In this regime, the corresponding $F$ matrix degenerates into a singular matrix. Therefore, we have assessed the performance of this detector by developing a new exact closed-form expression for the c.d.f. of the largest generalized eigenvalue of a complex correlated singular $F$-matrix when the underlying covariance assumes a rank-one spiked structure. An exact functional relationship between the detection probability and the false alarm rate (i.e., ROC profile) does not seem to be feasible for a general system configuration. However, when the noise-only sample covariance matrix is nearly rank deficient (i.e., $p=m$) such an explicit functional relationship has been obtained. This is one of the consequences of the powerful orthogonal polynomial approach that we have utilized in deriving the novel c.d.f.

It turns out that when $p=m$ the ROC profile corresponding to the largest sample generalized eigenvalue converges to the chance curve as $m\to\infty$, thereby proving that the sample generalized eigenvalue has no detection power in this regime. In sharp contrast to this observation, the same test statistic retains its power asymptotically should $\text{SNR}=O(m)$ for $m=p$. In this respect, we have obtained a remarkably simple asymptotic ROC profile, which as per our numerical investigations, serves as a good approximation to very small dimensional system configurations as well. This insightful observation is of paramount importance in detecting signals over Rayleigh fading channels for which the SNR scales with the system dimension $m$ for large $m$. 

Be that as it may, the question of further relaxing the restriction $p\geq m$ to accommodate the matrices with $p<m$ (i.e., the noise-only sample covariance matrix is also rank deficient) seems to be an important problem. Consequently, the noise-only sample covariance matrix becomes non-invertible and therefore, as per the literature,
 Moore-Penrose inverse can be used instead. In this respect, the distributional properties of the largest generalized eigenvalue under the null has been established in the literature. However, the statistical characterization under the alternative remains an important open problem. 
	\appendices
\section{Proof of the joint density}\label{appjoint}
Let us use \cite[Appendix A]{ref:AOSing} to rewrite the matrix integral in Theorem \ref{F_Def} as an integral over the unitary manifold to yield
\begin{align}
    f(\lambda_1,\cdots,\lambda_n) =\frac{2^n \pi ^{mn}\mathcal{K}_1(m,n,p)}{\tilde\Gamma_n(m)\det^{n}\left[\boldsymbol{\Sigma}\right]}
	\prod_{j=1}^n \lambda_j^{m-n}\Delta_n^2(\boldsymbol{\lambda})\int\limits_{\mathcal{U}_{m}} \frac{
  {\rm d}\mathbf{U}}{\det^{(n+p)}{\left[\mathbf{I}_m+\boldsymbol{\Sigma}^{-1}\mathbf{U} \tilde{\boldsymbol{\Lambda}}\mathbf{U}^\dagger\right]}}
\end{align}
where $\mathbf{U}=\left(\mathbf{U}_1\;\; \mathbf{U}_2\right)\in \mathcal{U}_m$ and $\tilde{\boldsymbol{\Lambda}}=\text{diag}\left(\lambda_1,\lambda_2,\ldots,\lambda_n,0,0,\ldots,0\right)\in\mathbb{R}^{m\times m}$. 
Following the decomposition of $\boldsymbol{\Sigma}$ given in (\ref{spike_invcov}), the above expression specializes to
\begin{align}
    f(\lambda_1,\cdots,\lambda_n) =\frac{2^n \pi ^{mn}\mathcal{K}_1(m,n,p)}{\tilde\Gamma_n(m)(1+\eta)^n}
	\prod_{j=1}^n \lambda_j^{m-n}\Delta_n^2(\boldsymbol{\lambda})\int\limits_{\mathcal{U}_{m}} \frac{
  {\rm d}\mathbf{U}}{\det^{(n+p)}{\left[\mathbf{I}_m+\mathbf{U} \tilde{\boldsymbol{\Lambda}}\mathbf{U}^\dagger-\mathbf{S}\boldsymbol{\Lambda}_\eta \mathbf{S}^\dagger\mathbf{U} \tilde{\boldsymbol{\Lambda}}\mathbf{U}^\dagger\right]}}
\end{align}
from which we obtain, nothing the factorization
\begin{align*}
    \det{\left[\mathbf{I}_m+\mathbf{U} \tilde{\boldsymbol{\Lambda}}\mathbf{U}^\dagger-\mathbf{S}\boldsymbol{\Lambda}_\eta \mathbf{S}^\dagger\mathbf{U} \tilde{\boldsymbol{\Lambda}}\mathbf{U}^\dagger\right]}&=\det\left[\mathbf{I}_m+\tilde{\boldsymbol{\Lambda}}\right]
    \det\left[\mathbf{I}_m-\mathbf{U}^\dagger\mathbf{S}\boldsymbol{\Lambda}_\eta \mathbf{S}^\dagger\mathbf{U} \tilde{\boldsymbol{\Lambda}}\left(\mathbf{I}_m+\tilde{\boldsymbol{\Lambda}}\right)^{-1}\right]\nonumber\\
    &=\prod_{j=1}^n (1+\lambda_j)  \det\left[\mathbf{I}_m-\boldsymbol{\Lambda}_\eta \mathbf{S}^\dagger\mathbf{U} \tilde{\boldsymbol{\Lambda}}\left(\mathbf{I}_m+\tilde{\boldsymbol{\Lambda}}\right)^{-1}\mathbf{U}^\dagger\mathbf{S}\right]  
\end{align*}
and $\mathbf{S}^\dagger\mathbf{U}\in\mathcal{U}_m$,
\begin{align}
    f(\lambda_1,\cdots,\lambda_n) &=\frac{2^n \pi ^{mn}\mathcal{K}_1(m,n,p)}{\tilde\Gamma_n(m)(1+\eta)^n}
	\prod_{j=1}^n \lambda_j^{m-n}(1+\lambda_j)^{-n}\Delta_n^2(\boldsymbol{\lambda})\nonumber\\
 & \qquad \qquad \times \int\limits_{\mathcal{U}_{m}} \frac{
  {\rm d}\mathbf{U}}{\det^{(n+p)}{\left[\mathbf{I}_m-\boldsymbol{\Lambda}_\eta \mathbf{U} \tilde{\boldsymbol{\Lambda}}\left(\mathbf{I}_m+\tilde{\boldsymbol{\Lambda}}\right)^{-1}\mathbf{U}^\dagger\right]}}.
\end{align}
Since $\boldsymbol{\Lambda}_\eta$ is a rank-one matrix, we may further simplify the above matrix integral to yield 
\begin{align}
\label{int_uni}
    f(\lambda_1,\cdots,\lambda_n) &=\frac{2^n \pi ^{mn}\mathcal{K}_1(m,n,p)}{\tilde\Gamma_n(m)(1+\eta)^n}
	\prod_{j=1}^n \lambda_j^{m-n}(1+\lambda_j)^{-n}\Delta_n^2(\boldsymbol{\lambda})\nonumber\\
 & \qquad \qquad \qquad \times \int\limits_{\mathcal{U}_{m}} \frac{
  {\rm d}\mathbf{U}}{{\left[1-\text{tr}\left(\boldsymbol{\Lambda}_\eta \mathbf{U} \tilde{\boldsymbol{\Lambda}}\left(\mathbf{I}_m+\tilde{\boldsymbol{\Lambda}}\right)^{-1}\mathbf{U}^\dagger\right)\right]^{n+p}}}.
\end{align}
Now keeping in mind that, for $\eta\geq 0$, 
\begin{align*}
   \text{tr}\left(\boldsymbol{\Lambda}_\eta \mathbf{U} \tilde{\boldsymbol{\Lambda}}\left(\mathbf{I}_m+\tilde{\boldsymbol{\Lambda}}\right)^{-1}\mathbf{U}^\dagger\right)&=\frac{\eta}{1+\eta}\mathbf{u}_1^\dagger \hspace{0.4mm}
   \text{diag}\left(\frac{\lambda_1}{1+\lambda_1},\ldots,\frac{\lambda_n}{1+\lambda_n},0,\ldots,0\right)\mathbf{u}\leq
   \frac{\eta}{(1+\eta)},
\end{align*}
we may invoke the integral relation
\begin{align}
    \frac{1}{s^{n+p}}=\frac{1}{\Gamma(n+p)}\int_0^\infty y^{n+p-1} \exp(-sy) {\rm d}y,\;\;\; \Re\{s\}>0
\end{align}
to further simplify (\ref{int_uni}) as
\begin{align}
\label{int_final}
    f(\lambda_1,\cdots,\lambda_n) &=\frac{2^n \pi ^{mn}\mathcal{K}_1(m,n,p)}{\tilde\Gamma_n(m)(1+\eta)^n}
	\prod_{j=1}^n \lambda_j^{m-n}(1+\lambda_j)^{-n}\Delta_n^2(\boldsymbol{\lambda})\int_0^\infty 
 y^{n+p-1} \exp(-y) \mathcal{I}(y) {\rm d}y
\end{align}
where 
\begin{align}
    \mathcal{I}(y)=\int\limits_{\mathcal{U}_m} \text{etr}\left(y\boldsymbol{\Lambda}_\eta \mathbf{U} \tilde{\boldsymbol{\Lambda}}\left(\mathbf{I}_m+\tilde{\boldsymbol{\Lambda}}\right)^{-1}\mathbf{U}^\dagger\right) {\rm d}\mathbf{U}
\end{align}
and we have changed the order of integration. Now to facilitate further analysis, noting that $\boldsymbol{\Lambda}_\eta$ is a rank-one matrix and $\tilde{\boldsymbol{\Lambda}}\left(\mathbf{I}_m+\tilde{\boldsymbol{\Lambda}}\right)^{-1}=\text{diag}\left(\frac{\lambda_1}{1+\lambda_1},\ldots,\frac{\lambda_n}{1+\lambda_n},0,\ldots,0\right)$ is rank deficient, we may use Theorem \ref{thmcontour} to rewrite the above matrix integral as
\begin{align}
  \mathcal{I}(y)=  \frac{\Gamma(m)}{2\pi \mathrm{i}} 
       \oint\limits_{\mathcal{C}}
        \frac{e^z}{z^{m-n}\displaystyle \prod_{j=1}^n \left(z- c_\eta\bar{\lambda}_j y\right)} {\rm d}z
\end{align}
where, for notation concision, we have used $\bar{\lambda}_j=\frac{\lambda_j}{1+\lambda_j},\; j=1,2,\ldots,n$.
Keeping in mind the $(m-n)$th order pole at the origin, we may evaluate the above contour integral to arrive at
\begin{align}
    \mathcal{I}(y)&=  \sum_{k=1}^n\frac{\Gamma(m)y^{1-m}\exp\left(c_\eta \bar{\lambda}_k y\right)}{c_\eta^{m-1}\bar{\lambda}_k^{m-n}\displaystyle \prod_{\substack{\ell=1\\ \ell \neq k}}^n\left(\bar{\lambda}_k-\bar{\lambda}_\ell\right)}+\sum_{k=1}^n\frac{\Gamma(m) y^{1-n}}{c_\eta^{n-1}\displaystyle \prod_{\substack{\ell=1\\ \ell \neq k}}^n\left(\bar{\lambda}_k-\bar{\lambda}_\ell\right)} \frac{1}{2\pi \mathrm{i}}
       \oint\limits_{\mathcal{O}}
        \frac{e^z}{z^{m-n}\displaystyle \left(z- c_\eta\bar{\lambda}_k y\right)} {\rm d}z\nonumber\\
        &=\sum_{k=1}^n\frac{\Gamma(m)y^{1-m}\exp\left(c_\eta \bar{\lambda}_k y\right)}{c_\eta^{m-1}\bar{\lambda}_k^{m-n}\displaystyle \prod_{\substack{\ell=1\\ \ell \neq k}}^n\left(\bar{\lambda}_k-\bar{\lambda}_\ell\right)}-\sum_{k=1}^n  \sum_{j=0}^{m-n-1} \frac{\Gamma(m) y^{-n-j}}
        {\Gamma(m-n-j)c_\eta^{n+j} \bar{\lambda}_k^{j+1}\displaystyle \prod_{\substack{\ell=1\\ \ell \neq k}}^n\left(\bar{\lambda}_k-\bar{\lambda}_\ell\right)}
\end{align}
where $\mathcal{O}$ is a small contour around the origin. Finally, we substitute $\mathcal{I}(y)$ into (\ref{int_final}) and perform term by term integration with some algebraic manipulations to conclude the proof.

		\section{Proof of the c.d.f. of the maximum eigenvalue }\label{appcdf}
	\noindent We find it convenient to derive the c.d.f. of the maximum of the transformed variables $y_j=\lambda_j/(1+\lambda_j),\; j=1,2,\ldots,n$, since the map $y\mapsto y/(y+1)$ preserves the order. To this end, following Corollary \ref{joint_eig_pdf}, we express the joint density of $y_1<y_2<\ldots<y_n<1$, after some algebraic manipulations, as
 \begin{align}
 \label{cdf_var_change}
  p(y_1,\ldots,y_n)&=f\left(\frac{y_1}{1-y_1},\ldots,\frac{y_n}{1-y_n}\right) \prod_{j=1}^n \frac{1}{(1-y_j)^2}\nonumber\\
  &=\frac{\mathcal{K}_2(m,n,p)}{(1+\eta)^n} \prod_{j=1}^n y_j^{m-n} (1-y_j)^{p-m} \Delta_n^2(\mathbf{y}) g(y_1,\ldots,y_n)
 \end{align}
 where $g(y_1,\ldots,y_n)$ is given by (\ref{eq_g}).
 Now by definition, the c.d.f. of $y_\max$ assumes the form
 \begin{align}
 \label{cdf_alt}
    \Pr(y_{\max}\leq x)=\idotsint\limits_{0<y_1<\ldots<y_n\leq x} p(y_1,\ldots,y_n) {\rm d}y_1 \ldots {\rm d}y_n
 \end{align}
 from which we obtain the desired c.d.f. as 
 \begin{align}
 \label{cdf trans}
      \Pr(\lambda_{\max}\leq x)= \Pr\left(y_{\max}\leq \frac{x}{1+x}\right).
 \end{align}
Therefore, our primary focus is on the evaluation of the multiple integral in (\ref{cdf_alt}). As such, noting that $p(y_1,\ldots,y_n)$ is symmetric in $y_1,\ldots,y_n$, we may remove the ordered region of integration to yield
\begin{align}
    \Pr(y_{\max}\leq x)=\frac{\mathcal{K}_2(m,n,p)}{(1+\eta)^n n!} &\int\limits_{(0,x)^n} 
    \sum_{k=1}^n
    \frac{1}{\displaystyle \prod_{\substack{\ell=1\\ \ell\neq k}}^{n}\left(y_k-y_\ell\right)}\left[\frac{\Gamma(n+p-m+1)}{c_\eta^{m-1} y_k^{m-n}\left(1-c_\eta y_k\right)^{n+p-m+1}}\right.\nonumber\\
    & \hspace{4.5cm}-\left.\sum_{j=0}^{m-n-1} \frac{\Gamma(p-j)}{\Gamma(m-n-j)c_\eta^{n+j}y_k^{j+1}}\right]\nonumber\\
    &\hspace{3.1cm}\times \prod_{i=1}^n y_i^{m-n} (1-y_i)^{p-m} \Delta_n^2(\mathbf{y}) {\rm d}y_1\ldots {\rm d}y_n. 
    \end{align}
Consequently, we can observe that each term in the above summation with respect to the index $k$ evaluates to the same value. Therefore, without loss of generality we may choose $k=1$ to further simplify the above multiple integral as 
\begin{align}
    \Pr(y_{\max}\leq x)=\frac{\mathcal{K}_2(m,n,p)}{(1+\eta)^n (n-1)!} &\int\limits_{(0,x)^n} 
    \frac{1}{\displaystyle \prod_{\ell=2}^{n}\left(y_1-y_\ell\right)}\left[\frac{\Gamma(n+p-m+1)}{c_\eta^{m-1} y_1^{m-n}\left(1-c_\eta y_1\right)^{n+p-m+1}}\right.\nonumber\\
    & \hspace{4.2cm}-\left.\sum_{k=0}^{m-n-1} \frac{\Gamma(p-k)}{\Gamma(m-n-k)c_\eta^{n+k}y_1^{k+1}}\right]\nonumber\\
    &\hspace{3.1cm}\times \prod_{i=1}^n y_i^{m-n} (1-y_i)^{p-m} \Delta_n^2(\mathbf{y}) {\rm d}y_1\ldots {\rm d}y_n. 
    \end{align}
Now we introduce the variable transformations $y_j=x s_j,\;\; j=1,\ldots,n$, which make the multi-dimensional region of integration independent of $x$, to obtain
\begin{align}
\label{eq_cdf_sub}
 \Pr(y_{\max}\leq x)= \mathcal{K}(m,n,p,\eta) x^{m(n-1)+1}\mathcal{A}(x)-\sum_{k=0}^{m-n-1}\mathcal{C}_k(m,n,p,\eta) x^{n(m-1)-k} \mathcal{B}_k(x)
\end{align}
where
\begin{align}
    \mathcal{A}(x)&=\int\limits_{(0,1)^n}\frac{\prod_{i=1}^n s_i^{m-n} (1-xs_i)^{p-m} \Delta_n^2(\mathbf{s})}
    {s_1^{m-n}\left(1-c_\eta x s_1\right)^{n+p-m+1}\prod_{\ell=2}^{n}\left(s_1-s_\ell\right)} {\rm d}s_1 \ldots {\rm d}s_n,\\
    \mathcal{B}_k(x)&=\int\limits_{(0,1)^n}\frac{\prod_{i=1}^n s_i^{m-n} (1-xs_i)^{p-m} \Delta_n^2(\mathbf{s})}
    {s_1^{k+1}\prod_{\ell=2}^{n}\left(s_1-s_\ell\right)} {\rm d}s_1 \ldots {\rm d}s_n, \label{Bdef}
\end{align}
\begin{align*}
    \mathcal{K}(m,n,p,\eta)&=\mathcal{K}_2(m,n,p)\frac{\Gamma(n+p-m+1)}{(n-1)! (1+\eta)^nc_\eta^{m-1}},
\end{align*}
and
\begin{align*}
    \mathcal{C}_k(m,n,p,\eta)&=\frac{\mathcal{K}_2(m,n,p)\Gamma(p-k)}{(n-1)!(1+\eta)^n\Gamma(m-n-k)c_\eta^{n+k}}.
\end{align*}

Let us now evaluate $\mathcal{A}(x)$ and $\mathcal{B}_k(x)$ both of which are structurally the same except a minor difference. Therefore, in what follows, we provide the details of the evaluation of $\mathcal{A}(x)$, whereas $\mathcal{B}_k(x)$ is evaluated in light of the former development. To be specific, we keep the integration with respect to $y_1$ last and rewrite $\mathcal{A}(x)$ in view of the decomposition $\Delta_n^2(\mathbf{s})=\prod_{i=2}^n (s_1-s_i)^2 \Delta_{n-1}^2(\mathbf{s})$ to arrive at
\begin{align}
\label{eq_A}
    \mathcal{A}(x)=\int_0^1 \frac{(1-xs_1)^{\alpha}}{\left(1-c_\eta x s_1\right)^{n+\alpha+1}}\mathcal{Q}_{n-1}(x,s_1){\rm d}s_1
\end{align}
where
\begin{align}
\label{Q}
    \mathcal{Q}_n(x,s_1)=\int\limits_{(0,1)^n} \Delta_{n}^2(\mathbf{z}) \prod_{j=1}^n
    z_j^{\beta}(1-xz_j)^{\alpha}(s_1-z_j) {\rm d}z_1\ldots {\rm d}z_n
\end{align}
with $\beta=m-n$.
The above multiple integral can be solved by leveraging the powerful orthogonal polynomial techniques advocated in \cite{me11}, as shown in Appendix \ref{appQeval}, to yield
\begin{align}
\label{eq_Qval}
    \mathcal{Q}_n(x,s_1)=K(n,\alpha,\beta)\frac{ x^{\alpha(n+1)}}{(1-s_1x)^{\alpha}}\det\left[P_{n+i-1}^{(0,\beta)}\left(2s_1-1\right)\hspace{5mm} \widetilde{\Psi}_{i,j}
(x)\right]_{\substack{i=1,2,...,\alpha+1\\j=2,3,...,\alpha+1}}
\end{align}
 where $\widetilde{\Psi}_{i,j}(x)=(n+i+\beta)_{j-2} P^{(j-2,\beta+j-2)}_{n+1+i-j}\left(\frac{2}{x}-1\right)$ and 
\begin{align*}
K(n,\alpha,\beta) &= \prod_{j=1}^{\alpha+1}\frac{(n+j-1)!(n+\beta+j-1)!}{(2n+2j+\beta-2)!} \prod_{j=0}^{n-1}\frac{j!(j+1)!(\beta+j)!}{(\beta+n+j)!}\prod_{j=0}^{\alpha-1}\frac{1}{j!}.
	\end{align*}
Having evaluated $\mathcal{Q}_n(x,s_1)$ in closed-form, we now substitute $\mathcal{Q}_{n-1}(x,s_1)$ into (\ref{eq_A}) with some cancellation of terms to obtain
\begin{align}
    \mathcal{A}(x)&=K(n-1,\alpha,\beta) x^{\alpha n}\int_0^1 \frac{1}{\left(1-c_\eta x s_1\right)^{n+\alpha+1}}\det\left[P_{n+i-2}^{(0,\beta)}\left(2s_1-1\right)\hspace{5mm} \Psi_{i,j}
(x)\right]_{\substack{i=1,2,...,\alpha+1\\j=2,3,...,\alpha+1}}{\rm d}s_1\nonumber\\
&=K(n-1,\alpha,\beta) x^{\alpha n}\det
\left[\int_0^1\frac{P_{n+i-2}^{(0,\beta)}\left(2s_1-1\right)}{\left(1-c_\eta x s_1\right)^{n+\alpha+1}} {\rm d}s_1\hspace{5mm} \Psi_{i,j}
(x)\right]_{\substack{i=1,2,...,\alpha+1\\j=2,3,...,\alpha+1}}\label{eq_Aint}
\end{align}
in which the second equality follows from the fact that only the first column of the determinant depends on $s_1$. Let us now focus on evaluating the inner integral in closed-form. To this end, following the definition (\ref{jacobidef}) of Jacobi polynomials, we may express the above integral as
\begin{align}
   \int_0^1\frac{P_{n+i-2}^{(0,\beta)}\left(2s_1-1\right)}{\left(1-c_\eta x s_1\right)^{n+\alpha+1}} {\rm d}s_1 &=
   \int_0^1\frac{{}_2F_1\left(-(n+i-2), n+\beta+i-2;1;1-s_1\right)}{\left(1-c_\eta x s_1\right)^{n+\alpha+1}}
    {\rm d}s_1\nonumber\\
  &=\frac{1}{\left(1-c_\eta x\right)^{n+\alpha+1}}
  \int_0^1\frac{{}_2F_1\left(-(n+i-2), n+\beta+i-2;1;t\right)}{\left(1+\displaystyle\frac{\eta x}{1+\eta(1-x)} t\right)^{n+\alpha+1}}
    {\rm d}t
\end{align}
where the second equality is due to the introduction of the variable transformation $t=1-s_1$. Since further manipulation in the current form is not possible, noting that the ${}_2F_1$ function assumes a finite series expansion given by
\begin{align}
    {}_2F_1\left(-(n+i-2), n+\beta+i-2;1;t\right)=\frac{(n+i-2)!}{(m+i-2)!}\sum_{k=0}^{n+i-2}
    \frac{(-1)^k(m+i+k-2)!}{(n+i-2-k)! k! k! } t^k
\end{align}
where we have used (\ref{apple}), the above integral can be simplified to give
\begin{align}
   \int_0^1\frac{P_{n+i-2}^{(0,\beta)}\left(2s_1-1\right)}{\left(1-c_\eta x s_1\right)^{n+\alpha+1}} {\rm d}s_1 
  =\frac{(n+i-2)!}{(m+i-2)!\left(1-c_\eta x\right)^{n+\alpha+1}} &
 \sum_{k=0}^{n+i-2}
    \frac{(-1)^k(m+i+k-2)!}{(n+i-2-k)! k! k! }\nonumber\\
    & \times \int_0^1\frac{t^k}{\left(1+\displaystyle\frac{\eta x}{1+\eta(1-x)} t\right)^{n+\alpha+1}}
    {\rm d}t
\end{align}
 which can be evaluated with the help of \cite[eq. 3.194]{ref:gradshteyn} to yield
\begin{align}
\label{eq_Pint}
   \int_0^1\frac{P_{n+i-2}^{(0,\beta)}\left(2s_1-1\right)}{\left(1-c_\eta x s_1\right)^{n+\alpha+1}} {\rm d}s_1 
  =\frac{\Omega^{(\alpha)}_{i}(x,\eta)}{\left(1-c_\eta x\right)^{n+\alpha+1}}.
\end{align}
We now substitute (\ref{eq_Pint}) into (\ref{eq_Aint}) and the resultant expression into (\ref{eq_cdf_sub}) followed by some algebraic manipulations to yield
\begin{align}
\label{cdf app}
     \Pr(y_{\max}\leq x)= & \frac{\mathcal{K}_\alpha(m,n)(n+\alpha)! x^{n(\alpha+m)-m+1}}{(1+\eta)^nc_\eta^{m-1}\left(1-c_\eta x\right)^{n+\alpha+1}}\det
\left[\Omega^{(\alpha)}_{i}(x,\eta)\hspace{5mm} \Psi_{i,j}
(x)\right]_{\substack{i=1,2,...,\alpha+1\\j=2,3,...,\alpha+1}}\nonumber\\
& \hspace{5cm} -\sum_{k=0}^{m-n-1}\mathcal{C}_k(m,n,p,\eta) x^{n(m-1)-k} \mathcal{B}_k(x).
\end{align}

The remaining task is to evaluate $\mathcal{B}_k(x)$. To this end, following the same lines of arguments as before, (\ref{Bdef}) can be written, keeping the integration with respect to $s_1$ last, as
\begin{align}
     \mathcal{B}_k(x)=\int_0^1 s_1^{\beta-k-1}(1-xs_1)^{\alpha}\mathcal{Q}_{n-1}(x,s_1){\rm d}s_1
\end{align}
 which can be evaluated using Appendix \ref{appQeval} to arrive at
 \begin{align}
    \mathcal{B}_k(x)&= K(n-1,\alpha,\beta) x^{\alpha n}\int_0^1 s_1^{\beta-k-1}\det\left[P_{n+i-2}^{(0,\beta)}\left(2s_1-1\right)\hspace{5mm} \Psi_{i,j}
(x)\right]_{\substack{i=1,2,...,\alpha+1\\j=2,3,...,\alpha+1}}{\rm d}s_1\nonumber\\
&=K(n-1,\alpha,\beta) x^{\alpha n}\det
\left[\int_0^1 s_1^{\beta-k-1}P_{n+i-2}^{(0,\beta)}\left(2s_1-1\right) {\rm d}s_1\hspace{5mm} \Psi_{i,j}
(x)\right]_{\substack{i=1,2,...,\alpha+1\\j=2,3,...,\alpha+1}}\label{eq_Bint}. 
 \end{align}
 The inner integral assumes, after the variable transformation $t=2s_1-1$, the following form
 \begin{align}
     \int_0^1 s_1^{\beta-k-1}P_{n+i-2}^{(0,\beta)}\left(2s_1-1\right) {\rm d}s_1 &=2^{k-\beta} \int_{-1}^1
     (1+t)^{\beta-k-1}P_{n+i-2}^{(0,\beta)}\left(t\right) {\rm d}t\nonumber\\
     &= (-1)^{n+i}\frac{(n+k+i-2)! (\beta-k-1)!}{k! (m+i-k-2)!}\label{B}
 \end{align}
 where the last equality follows from \cite[eq 7.391.2]{ref:gradshteyn} and ${}_2F_1(-n,b;c;1)=(c-b)_n/(c)_n$.
 Finally, we use (\ref{B}) in (\ref{eq_Bint}) and substitute the resultant expression into (\ref{cdf app}) followed by some algebraic manipulations with (\ref{cdf trans}) to conclude the proof.

\section{Proof of the c.d.f. of the maximum eigenvalue for $\eta=0$} \label{appcdfeta0}
Following the framework which leads to (\ref{cdf_var_change}), (\ref{cdf_alt}), (\ref{cdf trans}), and noting that 
\begin{align}
    p(y_1,\ldots,y_n)&=h\left(\frac{y_1}{1-y_1},\ldots,\frac{y_n}{1-y_n}\right)\prod_{j=1}^n \frac{1}{(1-y_j)^2}\nonumber\\
    &=\mathcal{K}_3(m,n,p) \prod_{j=1}^n y_j^{m-n}(1-y_j)^{p-m}  \Delta_n^2(\mathbf{y})
\end{align}
where $\mathcal{K}_3(m,n,p)=\frac{\pi^{n(n-1)}\widetilde{\Gamma}_m(n+p)}{\widetilde{\Gamma}_m(p)\widetilde{\Gamma}_n(n)\widetilde{\Gamma}_n(m) }$, 
we may write the corresponding c.d.f. as
\begin{align}
    \Pr\left(y_\max\leq x\right)=\mathcal{K}_3(m,n,p) \idotsint\limits_{0<y_1<\ldots<y_n\leq x} 
    \prod_{j=1}^n y_j^{\beta}(1-y_j)^{\alpha}  \Delta_n^2(\mathbf{y}) {\rm d}y_1\ldots {\rm d}y_n.
\end{align}
The ordered integration region can be made unordered regions due to symmetry as
\begin{align}
    \Pr\left(y_\max\leq x\right)=\frac{\mathcal{K}_3(m,n,p)}{n!} \int\limits_{(0,x)^n} 
    \prod_{j=1}^n y_j^{\beta}(1-y_j)^{\alpha}  \Delta_n^2(\mathbf{y}) {\rm d}y_1\ldots {\rm d}y_n
\end{align}
from which we obtain, keeping in mind that our focus is on using Jacobi polynomials, upon using the variable transformation $y_j=\frac{x}{2}(1+z_j),\;\; j=1,\ldots,n$,  
\begin{align}
    \Pr\left(y_\max\leq x\right)=\frac{\mathcal{K}_3(m,n,p)}{n! 2^{n(\alpha+m)}} x^{n(\alpha+m)}\int\limits_{(-1,1)^n} 
    \prod_{j=1}^n (1+z_j)^{\beta}\left(\omega_x-z_j\right)^{\alpha}  \Delta_n^2(\mathbf{z}) {\rm d}z_1\ldots {\rm d}z_n
\end{align}
where $\omega_x=\frac{2}{x}-1$.
Following [eqs. 22.1.10, 22.2.8]\cite{me11}, the above multiple integrals can be evaluated to yield
\begin{align}
    \Pr\left(y_\max\leq x\right)=\frac{\mathcal{K}_4(m,n,p)}{n! 2^{n(\alpha+m)}\prod_{j=0}^{\alpha-1} j!} 
    x^{n(\alpha+m)}\det\left[\frac{{\rm d}^\ell}{{\rm d}\omega_x^\ell}C_{n+k}\left(\omega_x\right)\right]_{k,\ell=0,1,...,\alpha-1}
\end{align}
where $C_k(x)$ are monic polynomials orthogonal with respect to the weight $(1+x)^\beta, \;\; x\in[-1,1]$ and $\mathcal{K}_4(m,n,p)=\mathcal{K}_3(m,n,p) 2^{n(\beta+n)}\prod_{j=0}^{n-1}\frac{j!(j+1)!(\beta+j)!}{(\beta+n+j)!}$. Clearly Jacobi polynomials satisfy the above requirement, therefore we select $ C_{k}(x) = 2^{k}\frac{k!(k+\beta)!}{(2k+\beta)!}P_{k}^{(0,\beta)}(x)$ and use (\ref{jacobiDerivative}) with some algebraic manipulations to obtain
\begin{align}
    \Pr\left(y_\max\leq x\right)=\frac{\mathcal{K}_4(m,n,p)}{n! 2^{nm}} & \prod_{j=0}^{\alpha-1}\frac{(m+j)!(n+j)!}{(m+n+2j)!j!}
    x^{n(\alpha+m)}\nonumber\\
    & \qquad \times \det\left[ (m+k+1)_{\ell}P_{n+k-\ell}^{(\ell,m-n+\ell)}
    \left(\frac{2}{x}-1\right)\right]_{k,\ell=0,1,...,\alpha-1}.
\end{align}
Finally, we use (\ref{cdf_alt}) and introduce the shift of indices $i=k+1,\, j=\ell+1$, followed by some algebraic manipulations to conclude the proof.

\section{Proof of Corollary \ref{corasy}}\label{app asym}
Let us further simplify (\ref{rocbalanced}) in light of  (\ref{eq_cdf0}), (\ref{eq_hyposimp}), and (\ref{apple}) to arrive at
\begin{align}
    P_D=1-
    S(m,\gamma,P_F)-(-1)^n
    T(m,\gamma,P_F)
\end{align}
where 
\begin{align}
    S(m,\gamma,P_F)&=n!\left(1+\frac{1}{\gamma}\right)^{m-n}\sum_{k=0}^{n-1}\sum_{\ell=0}^{n-k-1}
    \frac{(-1)^k(m+k-1)! }
    {(m-1)!k!(n-1-k-\ell)!(k+\ell+1)! \gamma^{n-\ell-1}}\nonumber\\
    &\hspace{6cm} \times \frac{\left(1-P_F\right)^{\frac{m(n-1)+\ell+1}{mn}}}{\left[1+\gamma\left(1-\left(1-P_F\right)^{1/mn}\right)\right]^{\ell+1}}
\end{align}
and
\begin{align}
\label{Tdef}
   T(m,\gamma,P_F)=\sum_{k=0}^{m-n-1}\frac{(n+k-1)!}{k!(n-1)! \gamma^n}\left(1+\frac{1}{\gamma}\right)^{k} \left(1-P_F\right)^{\frac{n(m-1)-k}{mn}}.
\end{align}
Our strategy is to evaluate the term-by-term limits of the above finite summations of functions as $m,\gamma\to\infty$ such that $\gamma/m\to c$. As such, we may write the desired limit in symbolic form as
\begin{align}
\label{asym symbolic}
    \lim_{\substack{m,\gamma\to\infty\\
    \frac{\gamma}{m}\to c}}  P_D=1-
     \lim_{\substack{m,\gamma\to\infty\\
    \frac{\gamma}{m}\to c}} S(m,\gamma,P_F)- \lim_{\substack{m,\gamma\to\infty\\
    \frac{\gamma}{m}\to c}} (-1)^n
    T(m,\gamma,P_F).
\end{align}
However, the $m$-dependent upper limit in the summation of $T(m,\gamma,P_F)$ creates a technical challenge in the direct evaluation of the corresponding limit. Therefore, let us evaluate the limits of $S(m,\gamma,P_F)$ and $T(m,\gamma,P_F)$ separately.

Let us focus on $S(m,\gamma,P_F)$. As such, noting the limits 
\begin{align}
\label{lim1}
    \lim_{\substack{m,\gamma\to\infty\\
    \frac{\gamma}{m}\to c}} \left(1+\frac{1}{\gamma}\right)^{m-n}=\exp\left(\frac{1}{c}\right)
\end{align}
and
\begin{align}
\label{lim2}
     \lim_{\substack{m,\gamma\to\infty\\
    \frac{\gamma}{m}\to c}} \gamma\left(1-\left(1-P_F\right)^{1/mn}\right)&=\lim_{\substack{m,\gamma\to\infty\\
    \frac{\gamma}{m}\to c}} \frac{\gamma}{m} \times \lim_{m\to\infty} m \left(1-\left(1-P_F\right)^{1/mn}\right)\nonumber\\
    &=-\frac{c}{n}\ln \left(1-P_F\right),
\end{align}
we may arrive at
\begin{align}
 \lim_{\substack{m,\gamma\to\infty\\
    \frac{\gamma}{m}\to c}}    S(m,\gamma,P_F)=n!\left(1-P_F\right)^{1-\frac{1}{n}}\exp\left(\frac{1}{c}\right) &
    \sum_{k=0}^{n-1}\sum_{\ell=0}^{n-k-1}
    \frac{(-1)^k \left(1-\frac{c}{n}\ln \left(1-P_F\right)\right)^{-\ell-1}}
    {k!(n-1-k-\ell)!(k+\ell+1)!}\nonumber\\
    & \qquad \times \frac{1}{c^{n-\ell-1}}
    \lim_{m\to\infty}  \frac{(m+k-1)!}{(m-1)!m^{n-\ell-1}}.
\end{align}
Following \cite[eq. 8.328.2]{ref:gradshteyn}, the remaining limit can be simplified to yield
\begin{align}
 \lim_{\substack{m,\gamma\to\infty\\
    \frac{\gamma}{m}\to c}}    S(m,\gamma,P_F)=n!\left(1-P_F\right)^{1-\frac{1}{n}}\exp\left(\frac{1}{c}\right) &
    \sum_{k=0}^{n-1}\sum_{\ell=0}^{n-k-1}
    \frac{(-1)^k \left[1-\frac{c}{n}\ln \left(1-P_F\right)\right]^{-\ell-1}}
    {k!(n-1-k-\ell)!(k+\ell+1)!}\nonumber\\
    & \qquad \qquad \times \frac{1}{c^{n-\ell-1}}
    \lim_{m\to\infty}  \frac{1}{m^{n-k-\ell-1}}.
\end{align}
Now it is worth observing that $ \displaystyle \lim_{m\to\infty}  \frac{1}{m^{n-k-\ell-1}}=0$ for $\ell<n-k-1$, whereas 
$ \displaystyle \lim_{m\to\infty}  \frac{1}{m^{n-k-\ell-1}}=1$ for $\ell=n-k-1$. Therefore, in the light of the preceding observation, we may obtain
\begin{align}
\label{Slimit}
     \lim_{\substack{m,\gamma\to\infty\\
    \frac{\gamma}{m}\to c}}    S(m,\gamma,P_F)=
    \left(1-P_F\right)^{1-\frac{1}{n}}\exp\left(\frac{1}{c}\right) &
    \sum_{k=0}^{n-1} \frac{(-1)^k}{k! c^k} \left[\frac{1}{1-\frac{c}{n}\ln \left(1-P_F\right)}\right]^{n-k}.
\end{align}
Since the above summation cannot be simplified further, we now focus on evaluating the limiting form of $T(m,\gamma,P_F)$.

To facilitate further analysis, let us rearrange the terms to rewrite (\ref{Tdef}) as
\begin{align}
\label{Tarrange}
   T(m,\gamma,P_F)= \frac{\left(1-P_F\right)^{\frac{n(m-1)}{mn}}}{(n-1)! \gamma^n}\sum_{k=0}^{m-n-1}\frac{(n+k-1)!}{k!}\left[\frac{1+\frac{1}{\gamma}}{\left(1-P_F\right)^{\frac{1}{mn}}}\right]^{k} 
\end{align}
to which the term-by-term limit taking process cannot be applied directly due to the $m$-dependent summation upper bound. To circumvent this difficulty, we make use of the following relationship\footnote{This can easily be proved by noting the identity
\begin{align*}
 \sum_{k=0}^{N-1} \frac{{\rm d}^{n-1}}{{\rm d}z^{n-1}} z^{n+k-1}=\frac{{\rm d}^{n-1}}{{\rm d}z^{n-1}}\left(\frac{z^{n-1}-z^{N+n-1}}{1-z}\right)=\sum_{k=0}^{n-1} \frac{(n-1)!}{k!(n-1-k)!} \frac{{\rm d}^{k}}{{\rm d}z^{k}}\left(\frac{1}{1-z}\right) \frac{{\rm d}^{n-1-k}}{{\rm d}z^{n-1-k}} \left(z^{n-1}-z^{N+n-1}\right)
\end{align*}
and the derivative relation $\frac{{\rm d}^{\ell}}{{\rm d}z^{\ell}} z^M=\frac{M!}{(M-\ell)!} z^{M-\ell},\;\; M\geq \ell$, where $M,\ell\in\mathbb{Z}^+\cup {0}$.}
\begin{align}
    \sum_{k=0}^{N-1} \frac{(n+k-1)!}{k!} z^k= \sum_{k=0}^{n-1}\frac{(n-1)!(n-1)! z^k}{k!(n-1-k)!(1-z)^{k+1}}-\sum_{k=0}^{n-1} \frac{(n-1)!(N+n-1)!  z^{N+k}}{(k+N)!(n-1-k)! (1-z)^{k+1}}
\end{align}
with $N=m-n$ and $z=\frac{1+\frac{1}{\gamma}}{\left(1-P_F\right)^{\frac{1}{mn}}}$ to rewrite (\ref{Tarrange}), after some algebraic manipulation, as
\begin{align}
T(m,\gamma,P_F)&= \sum_{k=0}^{n-1}
\frac{(-1)^{k+1}(n-1)!\left(1-P_F\right)^{\frac{n(m-1)+1}{mn}}\left(1+\frac{1}{\gamma}\right)^{k}}{k!(n-1-k)!\gamma^{n-k-1} \left[1+\gamma\left(1-\left(1-P_F\right)^{1/mn}\right)\right]^{k+1}}\nonumber\\
&  \quad- \sum_{k=0}^{n-1}
\frac{(-1)^{k+1}(m-1)!\left(1-P_F\right)^{\frac{m(n-1)+1}{mn}}\left(1+\frac{1}{\gamma}\right)^{m-n+k}}{(k+m-n)!(n-k-1)!\gamma^{n-1-k} \left[1+\gamma\left(1-\left(1-P_F\right)^{1/mn}\right)\right]^{k+1}}.
\end{align}
Since the above representation does not have $m$-dependent summation upper bound, we find it convenient to take term-by-term limits, keeping in mind the fundamental limiting expressions (\ref{lim1}) and (\ref{lim2}), to arrive at 
\begin{align}
     &\lim_{\substack{m,\gamma\to\infty\\
    \frac{\gamma}{m}\to c}}    T(m,\gamma,P_F)\nonumber\\
    &\quad =\sum_{k=0}^{n-1}
    \frac{(-1)^{k+1}(n-1)!\left(1-P_F\right)}{k!(n-1-k)!c^{n-k-1} 
    \left[1-\frac{c}{n}\ln\left(1-P_F\right)\right]^{k+1}}\times \lim_{m\to\infty}\frac{1}{m^{n-k-1}}\nonumber\\
    &\qquad\quad - \sum_{k=0}^{n-1}
\frac{(-1)^{k+1}\left(1-P_F\right)^{1-\frac{1}{n}}\exp\left(\frac{1}{c}\right)}{(n-k-1)!c^{n-1-k} \left[1-\frac{c}{n}\ln\left(1-P_F\right)\right]^{k+1}}\times 
\lim_{m\to\infty} \frac{(m-1)!}{(m+k-n)! m^{n-1-k}}.
\end{align}
Consequently we observe that the first sum is zero unless $k=n-1$ and the last limit evaluates to $1$ due to \cite[eq. 8.328.2]{ref:gradshteyn}. Therefore, the above limiting expression further simplifies to
\begin{align*}
    \lim_{\substack{m,\gamma\to\infty\\
    \frac{\gamma}{m}\to c}}    T(m,\gamma,P_F) &=\frac{(-1)^n\left(1-P_F\right)}{\left[1-\frac{c}{n}\ln\left(1-P_F\right)\right]^{n}}\\
    & \quad -
    \left(1-P_F\right)^{1-\frac{1}{n}}\exp\left(\frac{1}{c}\right)
    \sum_{k=0}^{n-1}
\frac{(-1)^{k+1}}{(n-k-1)!c^{n-1-k} \left[1-\frac{c}{n}\ln\left(1-P_F\right)\right]^{k+1}}
\end{align*}
from which we obtain, after introducing the index shift $\ell=n-1-k$,
\begin{align*}
    \lim_{\substack{m,\gamma\to\infty\\
    \frac{\gamma}{m}\to c}}    T(m,\gamma,P_F) &=\frac{(-1)^n\left(1-P_F\right)}{\left[1-\frac{c}{n}\ln\left(1-P_F\right)\right]^{n}}\\
    & \qquad -(-1)^n
    \left(1-P_F\right)^{1-\frac{1}{n}}\exp\left(\frac{1}{c}\right)
    \sum_{\ell=0}^{n-1}
\frac{(-1)^{\ell}}{\ell!c^{\ell}}\left[\frac{1}{1-\frac{c}{n}\ln\left(1-P_F\right)}\right]^{n-\ell}.
\end{align*}
Finally, in view of (\ref{Slimit}), we obtain
\begin{align*}
    \lim_{\substack{m,\gamma\to\infty\\
    \frac{\gamma}{m}\to c}}    S(m,\gamma,P_F) +(-1)^n\lim_{\substack{m,\gamma\to\infty\\
    \frac{\gamma}{m}\to c}}    T(m,\gamma,P_F)&=\frac{\left(1-P_F\right)}{\left[1-\frac{c}{n}\ln\left(1-P_F\right)\right]^{n}}
\end{align*}
which upon substituting into (\ref{asym symbolic}) gives the desired limit (\ref{cor asym limit}), thereby concluding the proof.

\section{The Evaluation of $\mathcal{Q}_n(x,t)$} \label{appQeval}
 Keeping in mind that the Jacobi polynomials are supported on the interval $[-1,1]$, let us change the region of integration in \eqref{Q} from $(0,1)^n$ to $(-1,1)^n$ by using the variable transformation  $ s_{j}=\frac{1+z_{j}}{2} $, $ j=1,2,...,n $, to yield 
	\begin{equation}\label{eqQapp}
	\mathcal{Q}_{n}(x,t) =  \frac{x^{\alpha n}}{2^{n(n+\beta+\alpha+1)}}\mathcal{R}_{n}(x,t) 
	\end{equation}
	where 
	\begin{align}
    \label{mehtaourint}
	\mathcal{R}_{n}(x,t) = \int\limits_{[-1,1]^{n}}^{} \prod_{j=1}^{n}(1+z_{j})^{\beta}&(\omega_x-z_{j})^{\alpha}\left(\omega_{\frac{1}{t}}-z_{j}\right)  \Delta_{n}^{2}(\mathbf{z}) {\rm d}z_1\ldots{\rm d}z_n,
	\end{align}
    with $ \omega_{x} = \frac{2}{x} -1$.
    Our strategy is to start with a related integral given in \cite[eqs. 22.4.2, 22.4.11]{me11} as
    \begin{align} \label{eqmehta}
		\int\limits_{[-1,1]^{n}}^{} \prod_{j=1}^{n}(1+z_{j})^{\beta}\prod_{i=1}^{\alpha+1}(r_{i}-z_{j}) \Delta_{n}^{2}(\mathbf{z}) {\rm d}z_1\ldots{\rm d}z_n                            
		=  \frac{K_{\beta,n}}{\Delta_{\alpha+1}(\mathbf{r})}\det\left[C_{n+i-1}(r_{j})\right]_{i,j=1,2,...,\alpha+1}
	\end{align}
    where
    \begin{align*}
    K_{\beta,n}=2^{n(\beta+n)}\prod_{j=0}^{n-1}\frac{j!(j+1)!(\beta+j)!}{(\beta+n+j)!}
    \end{align*}
    and $ C_{k}(x) $ are monic polynomials orthogonal with respect to the weight $ (1+x)^{\beta} $, over $ -1\leq x \leq 1 $. Since Jacobi polynomials are orthogonal with respect to the preceding weight, we use $ C_{k}(x) = 2^{k}\frac{k!(k+\beta)!}{(2k+\beta)!}P_{k}^{(0,\beta)}(x)$ in (\ref{eqmehta}) to obtain
    \begin{align}\label{eqmehta1}
	\int\limits_{[-1,1]^{n}} \prod_{j=1}^{n}(1+z_{j})^{\beta}  \prod_{i=1}^{\alpha+1}(r_{i}-z_{j}) \Delta_{n}^{2}(\mathbf{z}) {\rm d}z_1\cdots{\rm d}z_n
	 = \frac{\tilde{K}_{\beta,n}}{\Delta_{\alpha+1}(\textbf{r})}\det\left[P_{n+i-1}^{(0,\beta)}(r_{j})\right]_{i,j=1,2,...,\alpha+1}
	\end{align}
	where
	\begin{align*}
	\tilde{K}_{\beta,n} &= K_{\beta,n}\prod_{j=1}^{\alpha+1}\frac{2^{n+j-1}(m+j-1)!(n+\beta+j-1)!}{(2n+2j+\beta-2)!}.
	\end{align*}
  In the above, $r_i$s are generally distinct parameters. Nevertheless, if we choose $r_i$ such that
  \begin{align*}
  r_i=\left\{\begin{array}{ll}
  \omega_{\frac{1}{t}} & \text{if $i=1$}\\
  \omega_{x} & \text{if $i=2,3,\ldots,\alpha+1$},
  \end{array}\right.
  \end{align*}
    then the left side of (\ref{eqmehta1}) coincides with the multidimensional integral of our interest in (\ref{mehtaourint}). Under the above parameter selection, however, the right side of (\ref{eqmehta1}) takes the indeterminate form $0/0$. Therefore, to circumvent this difficulty, we have to evaluate following limit: 
    \begin{align}\label{khatrilimitq}
	&\mathcal{R}_{n}(x,t) = \tilde{K}_{\beta,n}\;\lim_{\substack{r_{1}\to \omega_{\frac{1}{x}}\\r_{2}, r_{3} ,...,r_{\alpha+1}\to \omega_{t}}}\frac{\det\left[P_{n+i-1}^{(0,\beta)}(r_{j})\right]_{i,j=1,2,...,\alpha+1}}{\Delta_{\alpha+1}(\mathbf{r})}.
	\end{align}
 To this end, following Khatri \cite{Khatri}, we write
 \begin{align}\label{eqkhatrilimit}
		&\lim_{\substack{r_{1}\to \omega_{\frac{1}{t}}\\r_{2}, r_{3} ,...,r_{\alpha+1}\to \omega_{x}}}\frac{\det\left[P_{n+i-1}^{(0,\beta)}(r_{j})\right]_{i,j=1,2,...,\alpha+1}}{\Delta_{\alpha+1}(\mathbf{r})}= 
        \frac{\det\left[P_{n+i-1}^{(0,\beta)}\left(\omega_{\frac{1}{t}}\right)\hspace{6mm} \frac{{\rm d}^{j-2}}{{\rm d}\omega_{x}^{j-2}}P_{n+i-1}^{(0,\beta)}(\omega_{x})\right]_{\substack{i=1,2,...,\alpha+1\\j=2,3,...,\alpha+1}}}
        {\det\left[\omega_{\frac{1}{t}}^{i-1}\hspace{6mm} \frac{{\rm d}^{j-2}}{{\rm d}\omega_{x}^{j-2}}\omega_{x}^{i-1}\right]_{\substack{i=1,2,...,\alpha+1\\j=2,3,...,\alpha+1}}}.
	\end{align}
    Now the determinant in the denominator of (\ref{eqkhatrilimit}) evaluates to
    \begin{align*}
    \det\left[\omega_{\frac{1}{t}}^{i-1}\hspace{6mm} \frac{{\rm d}^{j-2}}{{\rm d}\omega_{x}^{j-2}}\omega_{x}^{i-1}\right]_{\substack{i=1,2,...,\alpha+1\\j=2,3,...,\alpha+1}}=\prod_{j=1}^{\alpha-1}j!\left(\omega_{x}-\omega_{\frac{1}{t}}\right)^{\alpha}.
    \end{align*}
    The numerator can be rewritten with the help of (\ref{jacobiDerivative}) as
    \begin{align*}
    &\det\left[P_{n+i-1}^{(0,\beta)}\left(\omega_{\frac{1}{t}}\right)\hspace{6mm} \frac{{\rm d}^{j-2}}{{\rm d}\omega_{x}^{j-2}}P_{n+i-1}^{(0,\beta)}(\omega_{x})\right]_{\substack{i=1,2,...,\alpha+1\\j=2,3,...,\alpha+1}}\\
    &\qquad  =2^{-\frac{\alpha}{2}(\alpha-1)}\det\left[P_{n+i-1}^{(0,\beta)}\left(\omega_{\frac{1}{t}}\right) \qquad (n+\beta+i)_{j-2}P_{n+i-j+1}^{(j-2,\beta+j-2)}(\omega_{x})\right]_{\substack{i=1,2,...,\alpha+1\\j=2,3,...,\alpha+1}}.
    \end{align*}
  Substituting the above two expression into (\ref{eqkhatrilimit}) and then the result into (\ref{khatrilimitq}) gives
  \begin{align*}
		\mathcal{R}_{n}(x,t)&= \tilde{K}_{\beta,n}\frac{x^{\alpha}}{2^{\frac{\alpha}{2}(\alpha+1)}\prod_{j=1}^{\alpha-1}j!(1-xt)^{\alpha}}\\
		&\quad \qquad \qquad \times \det\left[P_{n+i-1}^{(0,\beta)}\left(\omega_{\frac{1}{t}}\right)\hspace{6mm}(n+i+\beta)_{j-2} P_{n+i-j+1}^{(j-2,\beta+j-2)}(\omega_{x})\right]_{\substack{i=1,2,...,\alpha+1\\j=2,3,...,\alpha+1}}
	\end{align*}	
	which upon substituting into (\ref{eqQapp}) followed by some algebraic manipulations  concludes the proof.

	\ifCLASSOPTIONcaptionsoff
	\newpage
	\fi

	
	
	%
\vspace{-1mm}
	\bibliographystyle{IEEEtran}
	\bibliography{bibfile}
	%
	
	
	
	
	
	
	

\end{document}